\documentclass[twocolumn,pra,letterpaper,aps,superscriptaddress,showpacs]{revtex4-1}

\usepackage{amsmath,upgreek,amssymb,graphicx,subfigure,dsfont,color,bbm}
\usepackage{soul}
\usepackage{enumitem}
\setitemize{leftmargin=*}
\usepackage{comment}
\usepackage{mathtools}
\usepackage[hidelinks]{hyperref}
\usepackage{xcolor}
\hypersetup{
    colorlinks,
    linkcolor={red!50!black},
    citecolor={blue!50!black},
    urlcolor={blue!80!black}
}
\usepackage[table]{xcolor}
\usepackage{csvsimple}
\usepackage{pgfplots}

\pgfplotsset{compat = newest}

\usepackage{lipsum}

\usepackage{fancyhdr}
\usepackage{tikz}

\usepackage[T1]{fontenc}
\usepackage{lmodern}
\usepackage{booktabs}% http://ctan.org/pkg/booktabs
\usepackage{multirow}
\usepackage[english]{babel}
\usepackage{fullpage}
\usepackage{verbatim}
\usepackage{float}
\usepackage{ bbold }
\usepackage{physics}
\usepackage{array}

\usepackage{scalerel}

\usepackage{natbib}
\usepackage{graphicx}
\usepackage{siunitx}
\usepackage{amssymb}
\usepackage{xspace}

\usepackage[utf8]{inputenc}

\graphicspath{{/home/victor/Escritorio/primerarticulo/Imagenes/}}

\DeclareMathAlphabet\mathbfcal{OMS}{cmsy}{b}{n}

\newcommand{\SigmaGauss}{\sigma_p}
\newcommand{\recoilfreq}{\omega_r}

\newcommand{\im}{\mathrm{i}}
\newcommand{\smalldet}{\delta}
\newcommand{\laserphase}{\Phi_\mathrm{L}}

\usepackage{color} 
\usepackage{titlesec}

\titleformat*{\subsubsection}{\bfseries}

\usepackage{natbib}
\begin{document}

\title{Diffraction-phase-free Bragg atom interferometry}

\author{V{\'{\i}}ctor Jos{\'{e}} Mart{\'{\i}}nez-Lahuerta}
\affiliation{Leibniz University Hannover, Institute of Quantum Optics, Hannover, Germany}

\author{Jan-Niclas Kirsten-Siem\ss}
\affiliation{Leibniz University Hannover, Institute of Quantum Optics, Hannover, Germany}

\author{Klemens Hammerer}
\affiliation{Leibniz University Hannover, Institute of Theoretical Physics, Hannover, Germany}
\affiliation{University Innsbruck, Institute of Theoretical Physics, Innsbruck, Austria}
\affiliation{Institute for Quantum Optics and Quantum Information, Austrian Academy of Sciences, 6020 Innsbruck, Austria}

\author{Naceur Gaaloul}
\affiliation{Leibniz University Hannover, Institute of Quantum Optics, Hannover, Germany}

\begin{abstract}

Bragg Diffraction of matter waves is an established technique used in the most accurate quantum sensors. It is also the method of choice to operate large-momentum-transfer, high-sensitivity atom interferometers. It suffers, however, from an intrinsic multi-path character. Optimal control theory (OCT) has recently led to an improved robustness of atom interferometers to a range of challenging environmental effects such as vibrations or platform accelerations. In this theoretical work, we apply OCT protocols to control the Bragg diffraction phase shifts thereby enhancing the metrological accuracy of the interferometer. We show a minimization of the diffraction phase for realistic conditions of finite temperature of the incoming wavepacket in a multi-path, high-order Bragg interferometer in a Mach-Zehnder configuration. We study input states with different momentum widths and find that our approach mitigates diffraction phases below the microradian level in the case of $1\%$ of the photon recoil, thereby eliminating one of the leading systematic effects in atom interferometry.

\end{abstract}

\date\today

\maketitle

%\tableofcontents

%%%%%%%%%%%%%%%%%%%%%%%%%%%%%%%%%%%%%%%%%%%%%%%%%%%%%%%%%%%%%%%%%%%%%%%%%%%
\section{Introduction}

Quantum sensors~\cite{QuantumSensingReview} possess a great sensitivity in probing small potential effects. They detect a great range of physical quantities such as magnetic and electric fields, time and frequency, acceleration and rotations, or temperature and pressure. To date, atom interferometers provide the most precise determination of the fine-structure constant~\cite{ParkerMeasurementfinestructureconstanttestStandardModel2018,Morel2020}, as well as the most accurate quantum test of the universality of free fall~\cite{AsenbaumAtomInterferometricTestEquivalencePrinciple2020}, by exploiting the interference of matter waves. Moreover, these sensors allow for absolute measurements of inertial forces with high accuracy and precision~\cite{Geiger2020}, making them ideally suited for practical applications~\cite{Bongs2019} such as gravimetry~\cite{Menoret2018,Wu2019_gravity}, gravity cartography~\cite{Stray2022}, and inertial navigation~\cite{Geiger2011,Cheiney-PRApp-2018}.
    
Great efforts are being made to dramatically increase the performance of state-of-the-art atom interferometers. On the one hand, this will aid in timely fundamental Physics quests such as the detection of gravitational waves or the probe for elusive phenomena such as dark matter and dark energy~\cite{Safronova2018}. On the other hand, active efforts are made to deploy real-world quantum sensors, e.g., through reduced integration times or more compact designs maintaining high sensitivity~\cite{Bongs2019}. This has been shown by pioneering experiments in the group of Ernst Rasel~\cite{AhlersPRL2016,Gebbe2021,AbendPRL2016}.

Bragg diffraction~\cite{Martin1988, Giltner1995} is one of the primary techniques used to enhance the sensitivity of atom interferometers through large momentum transfer~\cite{AsenbaumAtomInterferometricTestEquivalencePrinciple2020,AsenbaumPhaseShiftAtomInterferometerdueSpacetimeCurvatureitsWaveFunction2017,OverstreetObservationgravitationalAharonovBohmeffect2022, ParkerMeasurementfinestructureconstanttestStandardModel2018}. Bragg pulses impart multiple photon recoils onto atoms while the atom remains in its electronic ground state, allowing for state-of-the-art momentum separations~\cite{Chiow2011,PlotkinSwing2018,Gebbe2021,Wilkason2022,Rodzinka2024}. These pulses typically operate in the quasi-Bragg regime~\cite{EffectiveOmegaMuller} to balance scattering losses to parasitic states with the relatively strong velocity selectivity of the diffraction process~\cite{Szigeti_2012}. This compromise arises because Bragg scattering extends beyond a simple two-level system due to the relatively small energy splitting of the involved momentum states~\cite{EffectiveOmegaMuller, Siem2020}. As a result, the multi-path interference signal of a Bragg interferometer can deviate significantly from that of an idealized two-mode interferometer, especially for higher diffraction orders~\cite{Altin_AtomicGravimetryBragg2013,ParkerMultiportBraggdiffraction}.

These intrinsic multi-port and multi-path properties of Bragg interferometers cause diffraction phase shifts, which represent an important source of systematic errors~\cite{JanniPRL,ParkerMultiportBraggdiffraction,Beguin2022CharactBragg,PlotkinSwing2018,HighResolutionAIEstey2015,Bechner2003}. Recently, adapting the mirror pulse timing in the MZ configuration to reduce the populations of the parasitic paths was theoretically proposed~\cite{JanniPRL} and experimentally verified~\cite{Pfeifferdichroic2025}.
While this method has the potential to significantly suppress the diffraction phase in some geometries such as the MZ interferometer, its effectiveness appears limited when multiple parasitic diffraction orders are significantly populated, leaving the phase estimation of multi-port Bragg interferometers an open problem.

In this article, we use Optimal Control Theory (OCT) to engineer composite beam splitter (BS) and mirror (M) pulses that suppress the systematic phase shifts caused by the diffraction process to a high level. The effectiveness of using time-dependent interactions to maintain coherent control of complex quantum dynamics with high precision is well-known and has applications in quantum simulations, computation, and metrology~\cite{Souza2012}. Such techniques include shaped pulses~\cite{FREEMANNMRShapePulse,LuoYukunShapedRamanPulses2016,Fang_2018}, rapid adiabatic pulses~\cite{Baum1985AdiabaticInversion,KovachyAdiabaticRapidPassageBragg2012,BatemanAdiabaticPassage2LevelSystemMBSAI2007}, composite pulses~\cite{LEVITT1981CompositPulse,LEVITT1983CompositePulses2,Cummins_2000CompositeRotationsNMR,BergPAbendSRaselCompositeLigthPulse2015}, or continuous periodic fields~\cite{Fonseca_2005, Martínez-Lahuerta_2024,yalcinkaya_Continuous_2019, Bermudez_2012} all of which use time-dependent interactions to reproduce the equivalent of a robust single pulse. In the context of Bragg interferometry, the OCT has been shown to improve the diffraction efficiency and interferometer contrast of Bragg interferometers~\cite{LouieAI2023, QCTRLAI, Beguin2023, Li2024, baker2025robustquantumcontrolbragg}.

Here, we demonstrate the utility of OCT-engineered atom-light Bragg diffraction pulses in the metrology context by keeping the diffraction phase well below the mrad-level for a high-order Bragg interferometry scheme. Using the OCT to manipulate the Rabi frequency, phase, and detuning of these pulses, we mitigate the emergence of parasitic paths to ideally restore Bragg interferometry back to a two-mode operation.
Thus, we extend the use of OCT techniques by showcasing its capability to drive high-order Bragg transitions at vanishing diffraction phase, thereby advancing the practical applications of large-momentum-transfer atom interferometers.

In section~\ref{SectionBraggAI}, we introduce the effective multilevel Bragg Hamiltonian and discuss the experimentally relevant control parameters. We consider the diffraction of atomic states that feature a finite velocity distribution to account for Doppler effects. In section~\ref{SectionOCT}, we describe the OCT method and cost functions used to optimize the atom-light interaction. In section~\ref{PulseFidelity}, we assess the performance of individual pulses by comparing them to the unitary operations that would realize a two-mode MZ interferometer. We consider Bragg orders $n=3,5$ which are readily achievable in experiments in terms of laser power requirements and spontaneous emission~\cite{Siem2020,EffectiveOmegaMuller}. In section~\ref{AIFidelity}, we highlight the improved results using OCT pulses as contrasted to the more conventional pulses with Gaussian temporal shapes. In section~\ref{SectionDiffPhase}, we quantify the control of the diffraction phase in the MZ interferometers realized with OCT pulses. Finally, we summarize the findings of our study in section~\ref{SectionConclusions}.

\begin{figure}[h]
\centering
\subfigure[]{\includegraphics[width=0.49\linewidth]{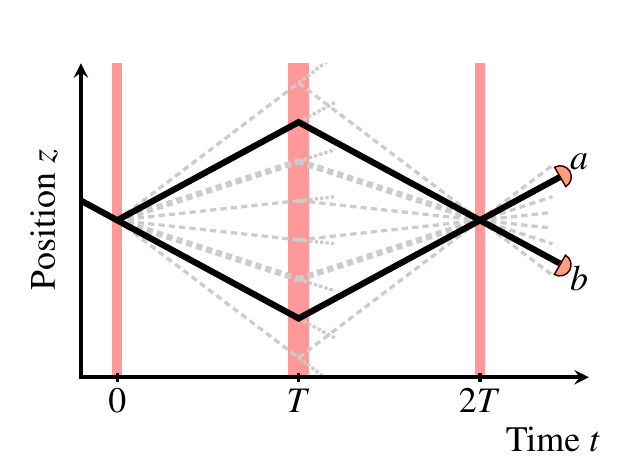}\label{Fig:MZI}}
\subfigure[]{\includegraphics[width=0.43\linewidth]{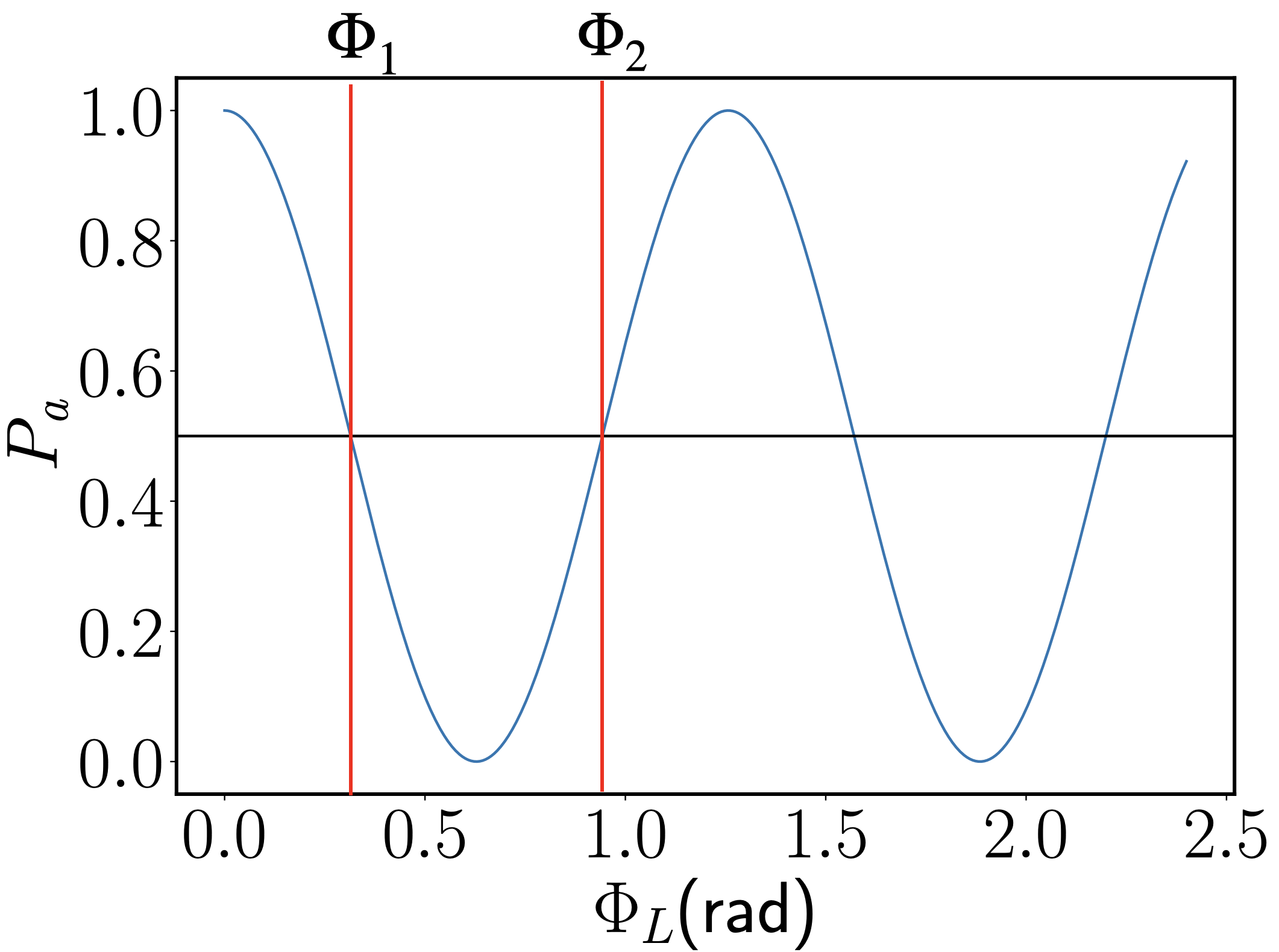}\label{Fig:MidfringesSketch}}
\caption{ a) Multi-path nature of Bragg atom interferometer for a MZ interferometer, showing an $n$th-order Bragg beam splitter populating the main trajectories (solid black lines), open ports, and parasitic paths (dashed lines; the dominant ones for $n=5$ are thicker), all affecting the Mach-Zehnder (MZ) signal recorded in ports a and b. Fig. adapted from~\cite{JanniPRL}. b) MZ signal phase scan of the population in port a ($P_a$) by scanning over the relative phase $\laserphase$ of the final beam splitter. The red lines represent the extraction of the mid-fringes by solving $P_a(\Phi)= 0.5$, where $P_a(\laserphase)$ is the fitted two-mode estimator for the population in port a, see Eq.~\eqref{eq:signalmodel}.}
\end{figure}

\section{High-order Bragg Atom Interferometry}\label{SectionBraggAI}

% The system 
%% Single-Bragg Bragg Hamiltonian

Bragg transitions are driven by elastic scattering from pulsed optical lattices. Two counter-propagating laser beams with wavevectors $\vec{k} = \vec{k}_1 \simeq -\vec{k}_2$ form an optical lattice, $V(t,x)=\Omega(t)\cos^2{\{  k[x-x_{\mathrm{L}}(t)]\}}$~\cite{P.Meystre.AtomOptics}. Typically, the lasers are far detuned from any electronically excited states and couple the atom's motional states with an effective two-photon Rabi frequency $\Omega(t)$~\cite{EffectiveOmegaMuller}. The lattice motion $x_{\mathrm{L}}(t) = \int^{t}_0 \smalldet (\tau)/2k\; \mathrm{d}\tau  + \laserphase(t)/k $ is a function of the laser frequency detuning $\smalldet (t) \equiv \omega_1(t) - \omega_2(t)$ and the relative laser phase $\laserphase(t) \equiv \phi_1(t)-\phi_2(t)$~\cite{Peik1997BlochOscillations}.

The pulsed Bragg lattice transfers multiple pairs of photon recoil $2n\hbar k$, where the integer $n$ is the Bragg order. When we expand the effective Hamiltonian in momentum eigenstates $\ket{\ell\hbar k }$ for integer $\ell$ and move to a frame co-moving with the optical lattice, we obtain the following matrix elements~\cite{LouieAI2023,EffectiveOmegaMuller}
\begin{align}\label{eq:BraggHamiltonian}
    H_{\ell,l}/\hbar =& [\ell^2\recoilfreq -\frac{\ell}{2}(4\recoilfreq \frac{p}{\hbar k}+\smalldet (t))]\delta_{\ell,l} \nonumber\\
    &+ \frac{\Omega(t)}{2}[e^{-\im\laserphase(t)}\delta_{\ell,l-2} + H.c.].
\end{align}
Here, $\recoilfreq \equiv \frac{\hbar k^2}{2M}$ is the recoil frequency of an atom with mass $M$. For a fixed even (odd) $n$ we only consider matrix elements with even (odd) $\ell$. In addition, we account for a momentum offset $p$ between the lattice and the atoms, along with the resulting Doppler effects~\cite{Szigeti_2012}. We assume that our initial atomic state has a Gaussian momentum distribution, which, without loss of generality, is initially centered around $\ket{-n \hbar k}$ and has a width $\SigmaGauss$, therefore, this means that the momentum offset  with respect to the center of the wavepacket, $p$, follows a normal distribution $\mathcal{N}(\mu,\sigma^2)$ with average $\mu=0$ and standard deviation $\sigma = \SigmaGauss$.

Both the intrinsic multipath nature of Bragg atom interferometry and the intrinsic momentum distribution of the incoming wave packet are accounted for by the Hamiltonian in Eq.~\eqref{eq:BraggHamiltonian}. While it describes, in principle, the coupling of an infinite ladder of momentum states, it is sufficient to consider a finite truncated Hilbert space due to finite coupling strengths~\cite{EffectiveOmegaMuller, Siem2020}. Here, we include $n+1$ discrete momentum states ($\ket{-n \hbar k}$, $\ket{-(n-2)\hbar k}$, $\ket{-(n-4) \hbar k}$,..., $\ket{(n-2) \hbar k}$,$\ket{n \hbar k}$) in addition to three additional levels at each end ($\ket{-(n + 6) \hbar k}$, $\ket{-(n+4) \hbar k}$, $\ket{-(n +2) \hbar k}$,$\ket{(n + 2) \hbar k}$,$\ket{(n+4) \hbar k}$, $\ket{(n+6) \hbar k}$). As there is no population in $\ket{\pm(n+6) \hbar k}$ states during the pulses, this extended space is sufficient to ensure convergence. Furthermore, three outer states fulfils the truncation criteria discussed in appendix B of Beguin \textit{et. al.}~\cite{Beguin2022CharactBragg}. Moreover, Eq.~\eqref{eq:BraggHamiltonian} reveals the control parameters to generate OCT pulses: The effective Rabi frequency, the relative laser phase, and the detuning: $\left(\Omega(t),\laserphase(t),\smalldet (t)\right)$. 

Although the relative laser phase $\Phi_L(t)$ and the frequency difference $\delta(t)$ are related at the level of the bare optical fields, after transformation to the partially comoving frame and adiabatic elimination, they enter the effective Hamiltonian in Eq.~\eqref{eq:BraggHamiltonian} through distinct operator structures. 
Specifically, $\delta(t)$ produces momentum-dependent diagonal shifts, whereas $\Phi_L(t)$ controls the complex phase of the off-diagonal momentum couplings. While a fully comoving frame allows the dynamics to be expressed in terms of a single lattice-position control (cf. Ref. \cite{Goerz2023}), this comes at the expense of coupling diagonal and off-diagonal control channels. For optimal control purposes, we therefore retain a partially comoving frame—analogous to Ref. \cite{QCTRLAI}—which provides greater control flexibility.

In contrast to previous studies, which were focused on improving the robustness of the interferometer against external noise sources~\cite{LouieAI2023,QCTRLAI,Li2024,baker2025robustquantumcontrolbragg}, we optimize these parameters to overcome the inherent multi-state scattering of Bragg diffraction and express the gains in terms of the metrologically relevant quantity: the diffraction phase.

\section{Pulse Optimizations}\label{SectionOCT}

Before evaluating the phase accuracy of the full OCT-enhanced Bragg interferometers, we optimize the individual beam-splitter and mirror operations making up the MZ in Fig.~\ref{Fig:MZI}. Here, we investigate how well they approximate effective two-mode operations in the presence of a significant velocity dispersion of the atomic wave packet. In fact, we observed that the approach of optimizing beam splitters and mirrors independently before assembling the full atom interferometer yields overall better results than optimizing the entire system at once.

% 1) Optimization parameters (Pulses): Gaussian, OCT
The OCT framework utilized in this study is Q-CTRL's Boulder Opal package~\cite{QCtrlpackage}. To quantify the improvements achieved by OCT optimization, we will compare their performance to pulses using optimized Gaussian temporal pulse shapes. It is, indeed, well-established in Bragg interferometry that Gaussian smooth flanks reduce scattering losses~\cite{EffectiveOmegaMuller, Siem2020} compared to box pulses.  
For the Gaussian pulses, we run the optimizer with a fixed laser phase and detuning $\laserphase(t) = 0 =\smalldet (t)$, while we set a Gaussian envelope for the effective Rabi frequency
\begin{align}\label{eq:GaussianPulse}
    \Omega(t) = \Omega_0 e^{-t^2/2\tau^2}.
\end{align}
This parametrization offers two optimization parameters: the peak Rabi frequency $\Omega_0$ and the pulse width $\tau$. The duration of the Gaussian pulse will be truncated at $\pm5\tau$.

In contrast, the optimization variables of the OCT pulses are $\Omega(t),\laserphase(t),\smalldet (t)$ with a total duration of $300\,\mu s$. To ensure that the optimized parameters remain within experimental constraints, we limit the maximum rate of change per time increment to $\Delta t= 1 \mu s$ for each parameter. To that end, we impose cut-off frequencies, $\omega_c$,  of $95$ kHz and $80$ kHz for beam splitters and mirrors, respectively. They take effect through a convolution of the optimization variables with a sinc kernel, $K(t)=\sin(\omega_c t)/\pi t$, defined by $\omega_c$. The sinc kernel in the frequency domain is constant in the range $[-\omega_c,\omega_c]$ and zero elsewhere.

In an experimental setting, it is common to only have access to the laser phase in order to control both the time-dependent detuning and the time-dependent phase. Therefore, to apply the OCT pulse with this constraint, the effective time-dependent phase generated by $\Phi_L^{exp}(t) = \Phi_L(t) + \int_0^t \delta(t^\prime)\, dt^\prime$, will have to be implemented, capturing the effects of both the time-dependent detuning and the time-dependent phase in a single effective time-dependent phase.

% 2) Optimization target + cost function
Ideally, the unitary time evolution operator for a two-mode beam splitter $\hat{U}_{\text{BS}}$ transfers the incoming state into an equal superposition, $\hat{U}_{\text{BS}}\ket{\mp n\hbar k}= \frac{1}{\sqrt{2}}\left(\ket{\mp n \hbar k}- \im \ket{\pm n \hbar k}\right)$. An ideal mirror reflects the momentum of the state, $\hat{U}_{\text{M}}\ket{\mp n\hbar k}= \ket{\pm n \hbar k}$. We will measure the performance of a given pulse by its distance to the target unitary $\hat{U}_T$, also referred to as gate fidelity, 
\begin{align}\label{eq:Fidelity}
    \mathcal{F}\left(\hat{U}_T\right) = \frac{1}{N}\sum_{i=1}^N \frac{\| \mathrm{Tr}\left(\hat{U}_T^{\dagger} \hat{U}^{(i)}\right) \|}{\| \mathrm{Tr}\left(\hat{U}_T^{\dagger} \hat{U}_T\right) \|}.
\end{align}

The cost function for the optimization is then the corresponding infidelity
\begin{align} \label{eq:avgInfidelity}
\text{cost}\left(\hat{U}_T\right) =\frac{1}{N} \sum_{i=1}^N 1 -\frac{1}{2}\| \mathrm{Tr}\left(\hat{U}_T^{\dagger} \hat{U}^{(i)}\right) \|.
\end{align}
In this equation, $\hat{U}_T$ are the respective target unitary operations, i.e., $\hat{U}_{\text{BS}}$ and $\hat{U}_{\text{M}}$. Robust control with respect to the finite momentum distribution of the incoming wave packet $\mathcal{N}(p,\SigmaGauss)$ is achieved as we average the cost for a batch of unitaries $\hat{U}^{(i)}$, which are characterized by their momentum offset $p_i$ with $i=1,\dots,N$ in Eq.~\eqref{eq:BraggHamiltonian}. For the results of the optimizations presented here, we consider wave packets with three different momentum widths $\SigmaGauss \in \{0.01,0.1,0.3\} \hbar k$, and sample them using batch sizes of $N=300$. To compare the performance of different OCT pulses, we increase the number of samples to $N=4000$. Finally, in Sec.~\ref{SectionDiffPhase}, in order to reach resolutions of the interferometer fringe of up to one $\mu$rad, we require on the order of $10^7$ samples for the Gaussian momentum distribution. 

Note, that we reduced the computational complexity of the optimization and evaluation of Eq.~\eqref{eq:avgInfidelity} by first projecting into the relevant two-mode subspace defined by the main momentum states $\ket{\mp n\hbar k}$~\cite{QCTRLAI}.
Moreover, we recall that in certain interferometer geometries like the MZ in Fig.~\ref{Fig:MZI}, the mirror pulse can serve a dual purpose by also deflecting incoming parasitic paths populated by the initial beam splitter~\cite{Pfeifferdichroic2025,JanniPRL}. As we will see, this is particularly important when using Gaussian pulse shapes, as the beam splitters exhibit non-negligible populations in parasitic paths. Consequently, we include the proper weights for deflecting the first-order parasitic paths in the cost function in the case of the Gaussian mirror pulses.

\section{Fidelity Analysis}\label{Results}
\subsection{Single Pulse}\label{PulseFidelity}

First, we evaluate the residual coupling to parasitic momentum states of both OCT pulses and Gaussian pulses using the optimized time-dependent pulses and parameters found by the optimizer. For all of the following results, we consider Bragg operations of orders $n=3,5$. Furthermore, third-order Bragg pulses couple to only a single pair of intermediate states compared to fifth-order pulses, suggesting a potential trade-off between Bragg order and parasitic effects.

\begin{figure}[h]
\subfigure[]{\includegraphics[width=0.49\linewidth]{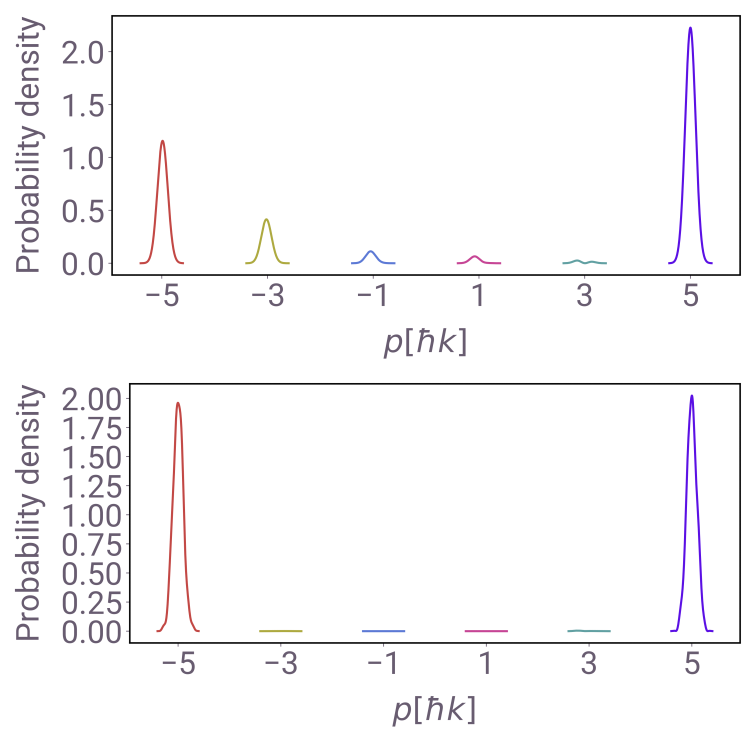}\label{Fig:ExampleBS}}
\subfigure[]{\includegraphics[width=0.49\linewidth]{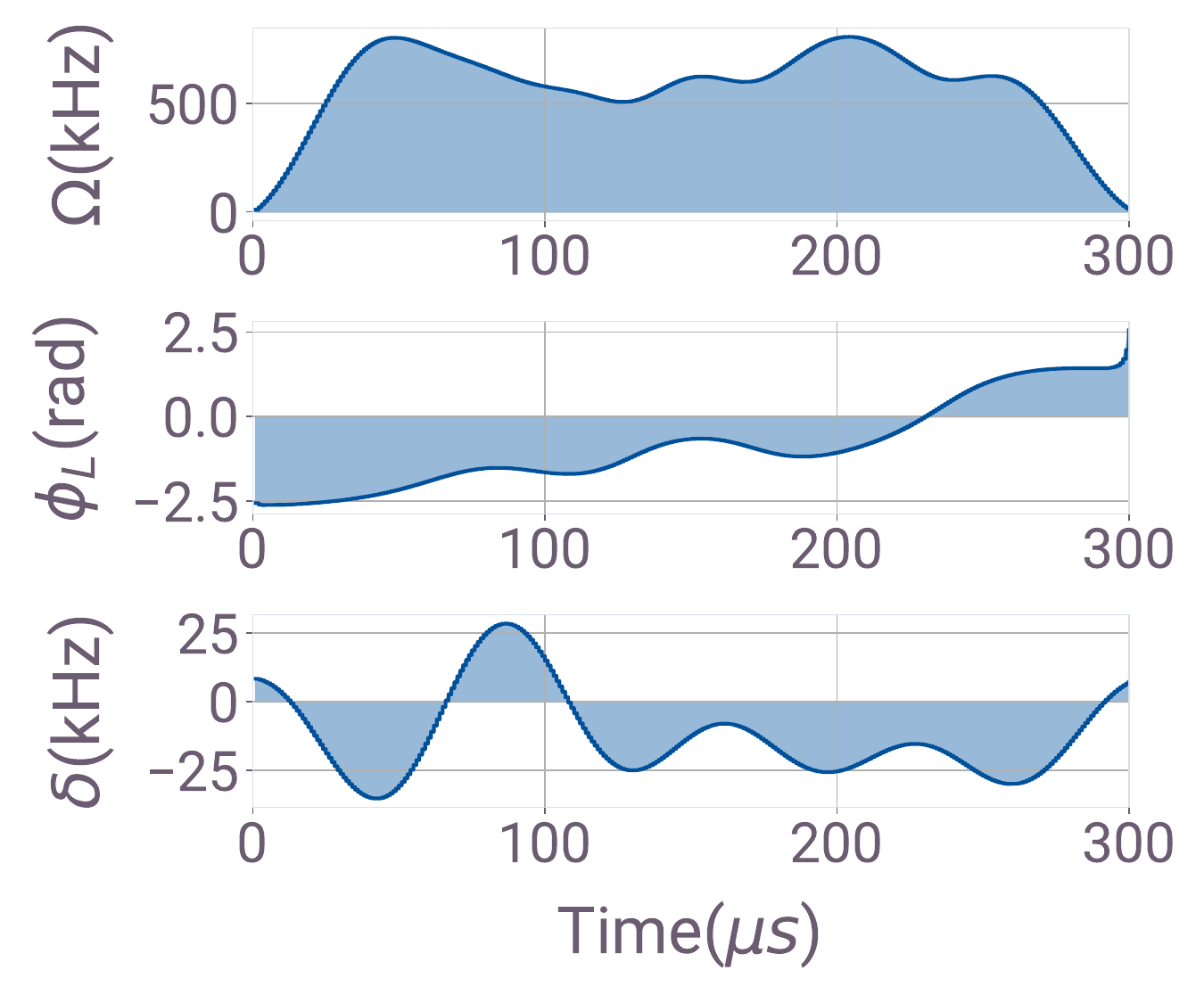}\label{Fig:ExampleBSParam}}
\caption{Study of an optimized beam splitter for the case of an initial state with momentum distribution of $\SigmaGauss = 0.1 \hbar k$ and Bragg order $n=5$. a) \textbf{Top}: Populations after the interaction with an optimal Gaussian pulse with a peak Rabi frequency of $\Omega_0 =34 \recoilfreq$ and a pulse width of $\tau=0.145 \recoilfreq^{-1}$, where we have the relative populations in $\ket{5\hbar k}$ and $\ket{-5\hbar k}$ of $0.5458$ and $0.2848$, respectively. \textbf{Bottom}: Populations after the interaction with an optimal OCT pulse where we have the relative populations in $\ket{5\hbar k}$ and $\ket{-5\hbar k}$ of $0.4997$ and $0.4980$, respectively b) OCT pulse parameters, where a cutoff frequency for the OCT of $\omega_c = 95$ kHz has been used for the optimization.}
\end{figure}

We start by discussing the population transfer of an incoming wave packet with momentum width $\SigmaGauss =0.1 \hbar k$, interacting with a Bragg beam splitter of order $n=5$ as depicted in Fig.~\ref{Fig:ExampleBS}. The plot shows the population distribution of the wave packet after the interaction with the beam splitter for both the Gaussian pulse parameters (Top) and OCT pulses (Bottom). As can be seen in Fig.~\ref{Fig:ExampleBS}(Top), the population in the momentum state $\ket{-3 \hbar k}$, which corresponds to one of the first parasitic paths, is about $10\%$ of the total population. This example illustrates clearly the challenge for traditional pulse shapes in achieving high fidelity while maintaining high diffraction efficiencies in the presence of considerable velocity dispersion and scattering losses. It also demonstrates that even for a narrow wave packet featuring $\SigmaGauss = 0.1 \hbar k$, the emerging parasitic populations are relevant. Note, that this is a result of using the cost function of Eq.~\eqref{eq:avgInfidelity} for optimizing the Gaussian pulses, same as in the OCT case. As we discuss in Appendix~\ref{App:DynPhases}, a different cost function optimizing Gaussian beam splitter efficiency rather than fidelity can lead to smaller parasitic populations. However, this comes at the price of  a higher phase variation over the momentum distribution as we explain in appendix~\ref{App:DynPhases}. As mentioned in the previous section, due to the fact that the population in the parasitic ports are not negligible in the Gaussian case, we have modified the cost function in Eq.~\eqref{eq:avgInfidelity} for the Gaussian mirror pulses according to the treatment of Appendix~\ref{App:CostGaussianMirror}.

In contrast, the OCT population transfer for the same incoming wave packet and Bragg order in Fig.~\ref{Fig:ExampleBS} (Bottom) shows an almost perfect beam splitter with smooth Gaussian distributions of the populations in $\ket{\pm 5 \hbar k}$ after the interaction. The corresponding pulse parameters for this OCT beam splitter are given in Fig.~\ref{Fig:ExampleBSParam}. In Appendix~\ref{App:populationcomparison}, we also show and discuss other combinations of $\SigmaGauss$ and OCT/Gaussian optimizations, as well as Bragg orders $n=3,5$, which illustrate that OCT may well suppress parasitic populations up to momentum widths of $\SigmaGauss = 0.3 \hbar k$. 

With this comparison, we see a significantly better suppression of parasitic paths using the OCT pulses compared to the Gaussian pulses, even at the level of the beam splitter, which will have a positive effect on the diffraction phase. Nevertheless, it is important to highlight that for $\SigmaGauss =0.3 \hbar k$, the parasitic ports of the OCT pulse do not disappear completely (see Appendix~\ref{App:populationcomparison}), and therefore colder clouds will have an advantage in terms of diffraction phase, as will be discussed in Sec.~\ref{SectionDiffPhase}.

% 5) More quantitativ results: Idealness

\subsection{Full Atom Interferometer}\label{AIFidelity}

The previous discussion of a single beam splitter has shown that OCT pulses can reach considerably higher population transfer for single operations, and  Table~\ref{Tab:n3n5} shows that they also reach higher fidelities especially for atomic clouds with significant velocity dispersion (cf. Refs. \cite{QCTRLAI} and \cite{LouieAI2023}). However, our focus is on suppressing the influence of parasitic paths on the signal of full interferometers. We recall that in these cases, later operations in the interferometer can be used to deflect parasitic paths~\cite{Pfeifferdichroic2025,JanniPRL}. Hence, we compute the fidelity of multiple consecutive operations, $\mathcal{F}\left(\hat{U}_T\right)$~\eqref{eq:Fidelity}, composing the MZ interferometer as shown in Fig.~\ref{Fig:MZI}. Here, $\hat{U}_T$ will correspond to the ideal beam splitter $\hat{U}_{BS}$, the ideal mirror $\hat{U}_{M}$, the ideal composite unitary after the mirror $\hat{U}_{M}\hat{U}_{BS}$ or the complete MZ unitary $\hat{U}_{BS}\hat{U}_{M}\hat{U}_{BS}$. Note, when calculating the fidelity for the composite unitary after the mirror and the complete MZ unitary, one must consider that momentum labels are not unique anymore, as multiple paths of the interferometer can have the same momentum. Therefore, one must account for this when defining these unitaries (See, for example, the supplement of Ref. \cite{JanniPRL}).

Table~\ref{Tab:n3n5} shows the fidelity of each step in the MZ interferometer. For the case of a small momentum distribution, $\SigmaGauss = 0.01 \hbar k$, the fidelity of the Gaussian pulses and the OCT pulses are comparable for both Bragg orders, even though the OCT pulses still perform better. Toward larger momentum widths, OCT starts to show a significant improvement compared to the Gaussian pulses, both for the individual operations and for the combined full interferometer ideal unitary. The most notable difference is observed for a Bragg order of $n=5$ and a momentum width of $\SigmaGauss = 0.1\hbar k$ where we have $\mathcal{F}= \{0.48,0.99\}$ for Gaussian and OCT-pulses, respectively. For the specific case of the mirror pulses, we can see that Gaussian pulses are capable of maintaining a high fidelity for $\SigmaGauss = 0.1 \hbar k$, but it drops considerably for $\SigmaGauss = 0.3 \hbar k$. Nevertheless, even for $\SigmaGauss = 0.1 \hbar k$ the full sequence is mainly determined by the performance of the beam splitter.

These results show how OCT pulses outperform Gaussian pulses if we want the interferometer to be as close as possible to a two-mode interferometer, i.e., achieving better suppression of parasitic paths and less sensitivity to velocity dispersion. Furthermore, we also observe that colder clouds exhibit considerably better performance. To clarify what that means in terms of accuracy, we will now compute the diffraction phase of these interferometers.

\newcommand{\getColumnValue}[2]{% #1 = row number, #2 = column name
    \csvreader[head to column names]{Newtabledata5U.csv}{}{%
        \ifnum\thecsvrow=#1 % Check if it's the x-th row
            \pgfmathparse{#2} % Parse the value of the specified column
            \def\ColumnValue{\pgfmathresult} % Store the value in the macro
        \fi
    }%
}
\newcommand{\getColumnValues}[2]{% #1 = row number, #2 = column name
    \csvreader[head to column names]{Newtabledata3U.csv}{}{%
        \ifnum\thecsvrow=#1 % Check if it's the x-th row
            \pgfmathparse{#2} % Parse the value of the specified column
            \def\ColumnValue{\pgfmathresult} % Store the value in the macro
        \fi
    }%
}

\begin{table}[h!]
\centering
\begin{tabular}{c c >{\columncolor{gray!15}}c >{\columncolor{blue!8}}c >{\columncolor{gray!15}}c >{\columncolor{blue!8}}c}
%\hline
&&\multicolumn{2}{c}{$n=3$}&\multicolumn{2}{c}{$n=5$}\\
\multirow{1}{*}{}  & $\SigmaGauss\;[\hbar k]$ & Gaussian & OCT & Gaussian & OCT \\
\midrule
& $0.01$ & \getColumnValues{4}{\BS}\ColumnValue& \getColumnValues{1}{\BS }\ColumnValue & \getColumnValue{4}{\BS}\ColumnValue & \getColumnValue{1}{\BS }\ColumnValue \\
$\hat{U}_\mathrm{BS}$ & $0.1$ & \getColumnValues{5}{\BS}\ColumnValue & \getColumnValues{2}{\BS} \ColumnValue & \getColumnValue{5}{\BS}\ColumnValue & \getColumnValue{2}{\BS}\ColumnValue \\
& $0.3$ & \getColumnValues{6}{\BS}\ColumnValue & \getColumnValues{3}{\BS}\ColumnValue & \getColumnValue{6}{\BS}\ColumnValue & \getColumnValue{3}{\BS}\ColumnValue \\  
\midrule
& $0.01$ & \getColumnValues{4}{\M}\ColumnValue& \getColumnValues{1}{\M }\ColumnValue & \getColumnValue{4}{\M}\ColumnValue & \getColumnValue{1}{\M }\ColumnValue \\
$\hat{U}_\mathrm{M}$ & $0.1$ & \getColumnValues{5}{\M}\ColumnValue & \getColumnValues{2}{\M} \ColumnValue & \getColumnValue{5}{\M}\ColumnValue & \getColumnValue{2}{\M}\ColumnValue \\
& $0.3$ & \getColumnValues{6}{\M}\ColumnValue & \getColumnValues{3}{\M}\ColumnValue & \getColumnValue{6}{\M}\ColumnValue & \getColumnValue{3}{\M}\ColumnValue \\  
\midrule
& $0.01$ & \getColumnValues{4}{\MBS}\ColumnValue& \getColumnValues{1}{\MBS }\ColumnValue & \getColumnValue{4}{\MBS}\ColumnValue & \getColumnValue{1}{\MBS }\ColumnValue \\
$\hat{U}_\mathrm{M}\hat{U}_\mathrm{BS}$ & $0.1$ & \getColumnValues{5}{\MBS}\ColumnValue & \getColumnValues{2}{\MBS} \ColumnValue & \getColumnValue{5}{\MBS}\ColumnValue & \getColumnValue{2}{\MBS}\ColumnValue \\
& $0.3$ & \getColumnValues{6}{\MBS}\ColumnValue & \getColumnValues{3}{\MBS}\ColumnValue & \getColumnValue{6}{\MBS}\ColumnValue & \getColumnValue{3}{\MBS}\ColumnValue \\  
\midrule
& $0.01$ & \getColumnValues{4}{\BSMBS}\ColumnValue& \getColumnValues{1}{\BSMBS }\ColumnValue & \getColumnValue{4}{\BSMBS}\ColumnValue & \getColumnValue{1}{\BSMBS }\ColumnValue \\
$\hat{U}_\mathrm{BS}\hat{U}_\mathrm{M}\hat{U}_\mathrm{BS}$ & $0.1$ & \getColumnValues{5}{\BSMBS}\ColumnValue & \getColumnValues{2}{\BSMBS} \ColumnValue & \getColumnValue{5}{\BSMBS}\ColumnValue & \getColumnValue{2}{\BSMBS}\ColumnValue \\
& $0.3$ & \getColumnValues{6}{\BSMBS}\ColumnValue & \getColumnValues{3}{\BSMBS}\ColumnValue & \getColumnValue{6}{\BSMBS}\ColumnValue & \getColumnValue{3}{\BSMBS}\ColumnValue \\  
\bottomrule
\end{tabular}
\caption{Fidelity, as defined by Eq.~\eqref{eq:Fidelity}, of a beam splitter, a mirror, their combination, or the full interferometer for optimal control pulses and optimized Gaussian pulses for Bragg orders $n=3$ and $n=5$ and an initial state with a momentum Gaussian distribution of width $\SigmaGauss$ centered at $\ket{-3\hbar k}$ and $\ket{-5\hbar k}$ respectively. The background colors (gray for Gaussian and blue for OCT) are given to facilitate the reading.  }
\label{Tab:n3n5}
\end{table}

\section{Phase accuracy}\label{SectionDiffPhase}
%Transition and definition of diffraction phase
While fidelity provides an efficient and well-defined objective for optimal pulse design, the performance of the resulting OCT MZ interferometer is ultimately evaluated in terms of the phase accuracy using experimentally relevant observables, namely contrast and diffraction-induced phase shifts. The diffraction-induced phase shifts are quantified by computing the residual phase shift caused by the spurious couplings to unwanted momentum states, i.e., the Bragg diffraction phase~\cite{JanniPRL,ParkerMultiportBraggdiffraction,Beguin2022CharactBragg,Geiger2020,PlotkinSwing2018,HighResolutionAIEstey2015,Bechner2003}. In the absence of other phase contributions, we compute the diffraction phase, denoted as $\delta \Phi \equiv \Phi^{\text{measured}}-\laserphase$, by subtracting the control phase shift $\laserphase$, e.g., imprinted via the laser phase of the final pulse of the MZ, from the phase extracted from the interferometer signal $\Phi^{\text{measured}}$.   
%Measurement and phase extraction procedure
Atom interferometers encode the relative phase between the arms of the interferometer $\Phi$ in the populations of their output ports. The interferometer signal is typically defined as the relative atom numbers measured in the two main ports, e.g., $N_a$ and $N_b$ in Fig.~\ref{Fig:MZI},
\begin{align}
    P_a(\Phi) = \frac{N_a(\Phi)}{N_a(\Phi)+N_b(\Phi)}.\label{eq:signalpop}
\end{align}
To extract the phase from population measurements, one must define an estimator that describes the functional relationship between these two quantities. In the case of ideal two-mode beam splitters and mirrors, this signal is a perfect cosine, $P_a(\Phi)\propto \cos{(\Phi)}$. This motivates the widespread use of the two-mode estimator 
\begin{align}\label{eq:signalmodel}
    P_a(\Phi) = \frac{1}{2}\left(B + C \cos(n\Phi + A)\right),
\end{align}
where $P_b(\Phi)$ simply includes an offset of $\pi$ in the phase. The fit parameters in this formula are a shift in the phase $A$, an offset in the mean amplitude of the signal $B$ and its amplitude $C$.
Varying the laser phase shift $\laserphase$ produces an oscillating signal as depicted in Fig.~\ref{Fig:MidfringesSketch}, to which the model in Eq.~\eqref{eq:signalmodel} is fit, calibrating the parameters $A,B$ and $C$. Finally, the estimated phase is a function of the measured populations and can be obtained by inverting the calibrated signal model $\Phi^{\text{measured}} = {P^{-1}}_a(P^{\text{measured}}_a)$. Yet, the scattering to spurious states gives rise to a more complex signal in Bragg atom interferometers than the one in Eq.~\eqref{eq:signalmodel} ~\cite{Altin_AtomicGravimetryBragg2013,JanniPRL, ParkerMultiportBraggdiffraction}. Applying a more complex estimator model, which accounts for all possible Fourier components in the interferometer signal, would be challenging to implement in an experiment. This is due to the large number of parameters involved and the difficulty of detecting a very small number of atoms per port. Implementing the model in~\cite{JanniPRL}, which accounts for a few Fourier components, would assume a mirror interaction transparent for the main parasitic paths. Instead of these approaches, we proceed to quantify the residual errors in terms of $\delta \Phi$ when using the OCT pulses found in section~\ref{Results}. 

\begin{widetext}

\begin{figure}[H]
\centering
% Column titles
\begin{minipage}[b]{0.05\textwidth}
\hfill
\end{minipage}
\begin{minipage}[b]{0.3\textwidth}
\centering
\textbf{$\SigmaGauss = 0.01 \hbar k$}
\end{minipage}
\begin{minipage}[b]{0.3\textwidth}
\centering
\textbf{$\SigmaGauss = 0.1 \hbar k$}
\end{minipage}
\begin{minipage}[b]{0.3\textwidth}
\centering
\textbf{$\SigmaGauss = 0.3 \hbar k$}
\end{minipage}

\vspace{0.2cm}

% First row title and figures
\begin{minipage}[b]{0.05\textwidth}
\centering
\raisebox{2cm}{\textbf{$\Phi_1$}} % Adjust the raisebox value to center vertically
\end{minipage}
\begin{minipage}[b]{0.3\textwidth}
\centering
\subfigure{\includegraphics[width=\textwidth]{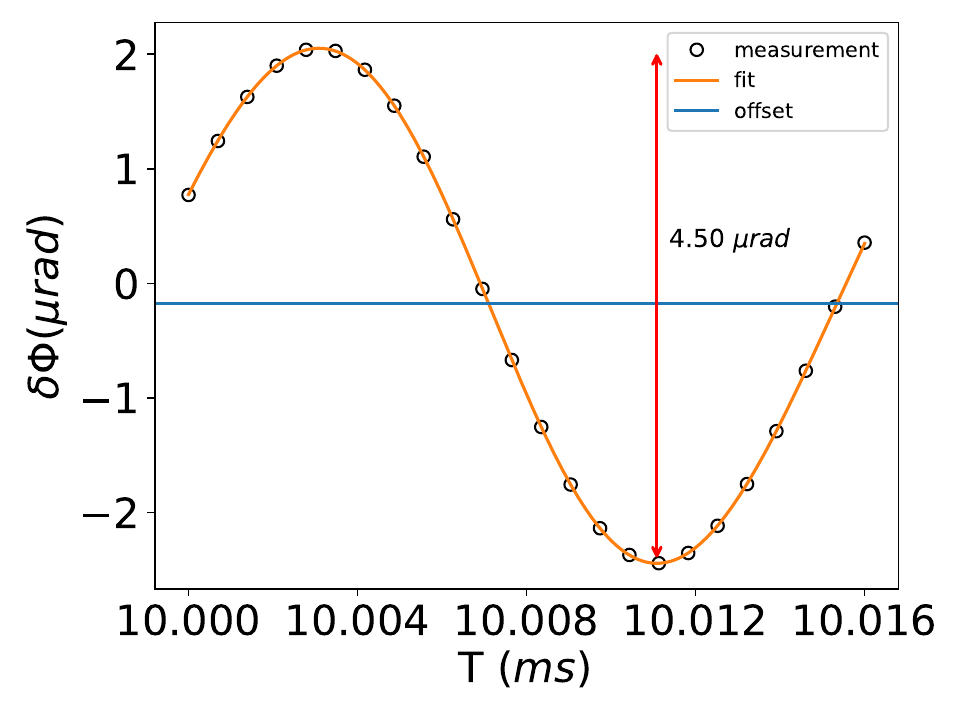}}
\end{minipage}
\begin{minipage}[b]{0.3\textwidth}
\centering
\subfigure{\includegraphics[width=\textwidth]{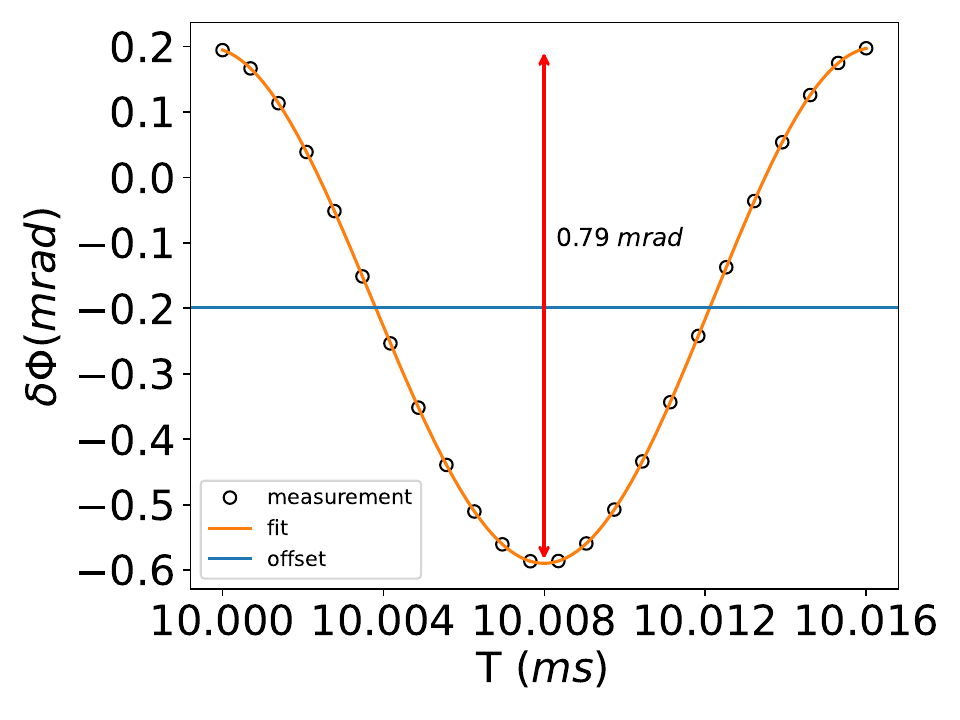}}
\end{minipage}
\begin{minipage}[b]{0.3\textwidth}
\centering
\subfigure{\includegraphics[width=\textwidth]{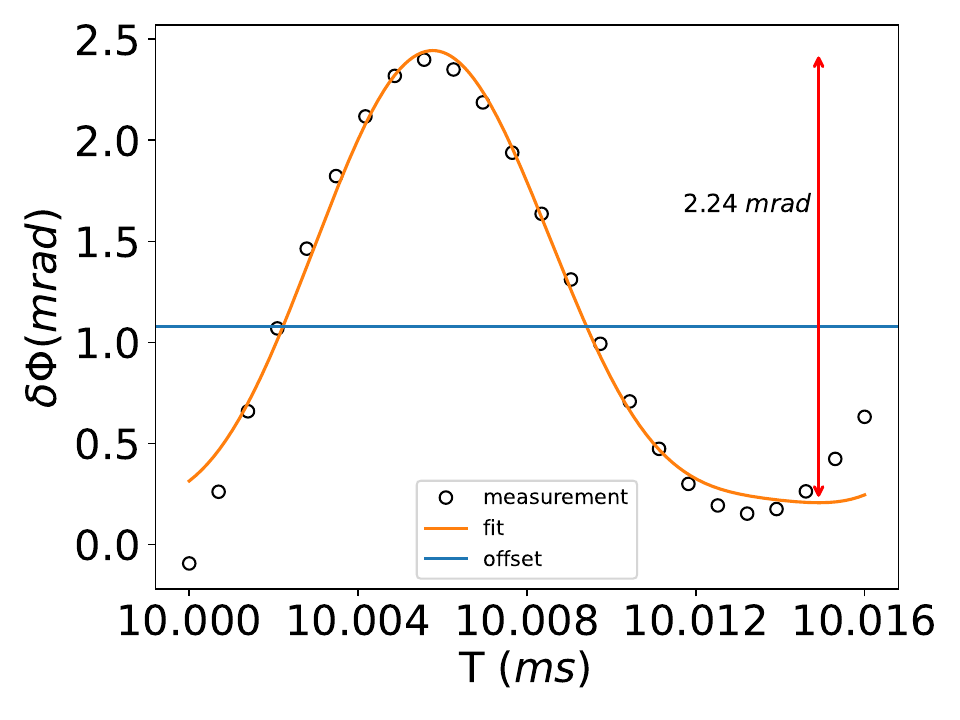}}
\end{minipage}

\vspace{0.2cm}

% Second row title and figures
\begin{minipage}[b]{0.05\textwidth}
\centering
\raisebox{2cm}{\textbf{$\Phi_2$}} % Adjust the raisebox value to center vertically
\end{minipage}
\begin{minipage}[b]{0.3\textwidth}
\centering
\subfigure{\includegraphics[width=\textwidth]{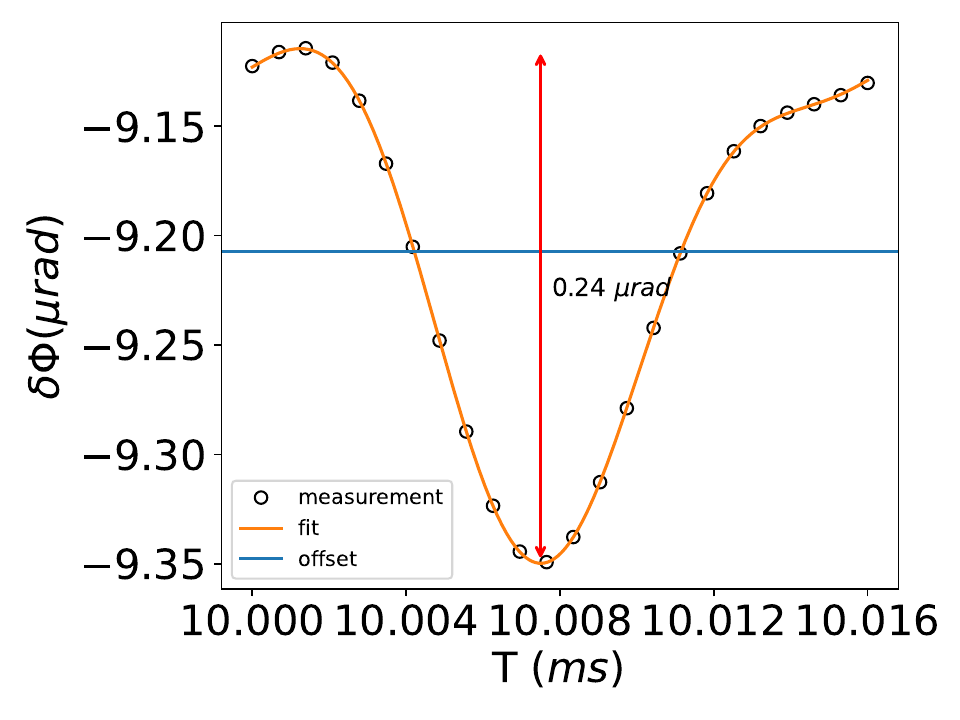}}
\end{minipage}
\begin{minipage}[b]{0.3\textwidth}
\centering
\subfigure{\includegraphics[width=\textwidth]{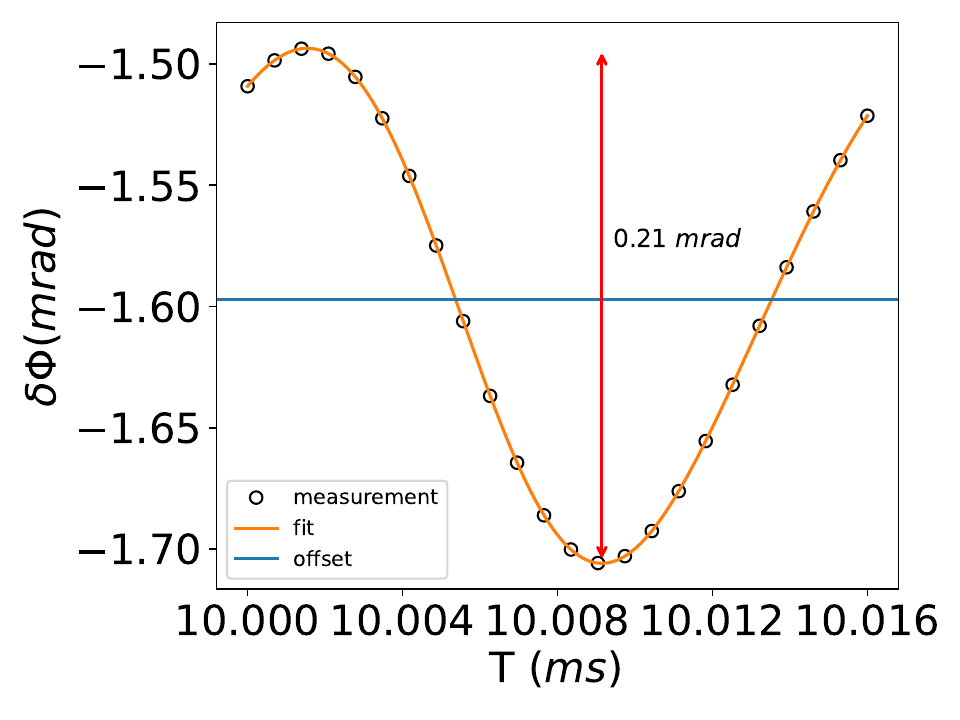}}
\end{minipage}
\begin{minipage}[b]{0.3\textwidth}
\centering
\subfigure{\includegraphics[width=\textwidth]{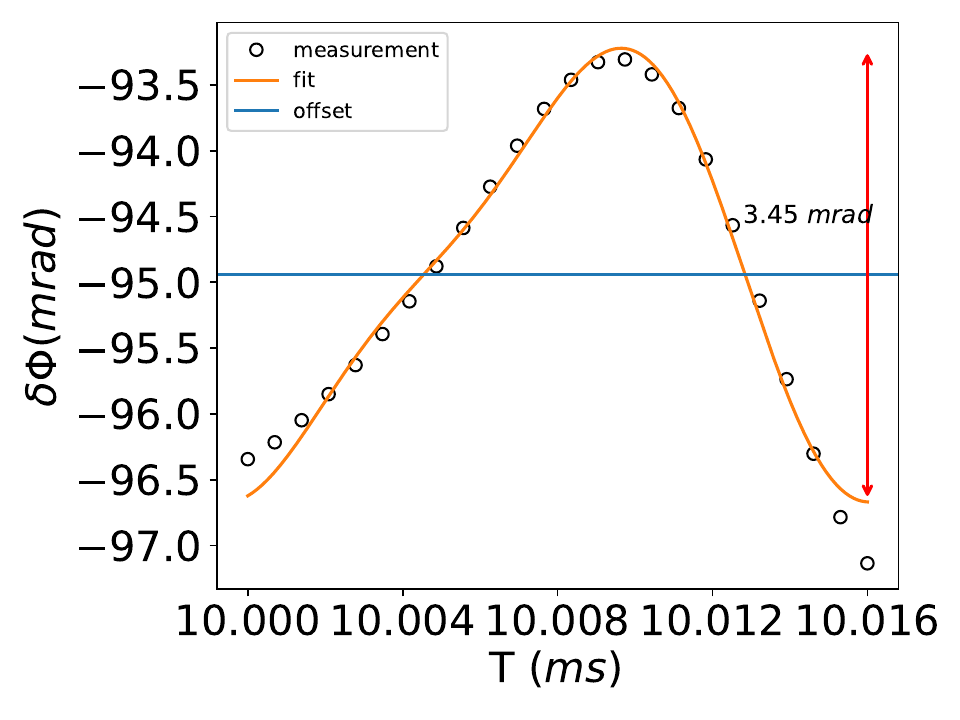}}
\end{minipage}

\caption{Diffraction phase for Bragg diffraction order $n=3$. The first and second rows show the diffraction phase at the first and second mid-fringe, respectively. The columns, from left to right, correspond to $\SigmaGauss \in \{0.01,0.1,0.3\}\,\hbar k$. In each plot, the solid blue horizontal line indicates the systematic phase shift, the empty circles represent the simulated data, and the orange curve corresponds to Eq.~\eqref{eq:OscFitModel} fitted to the simulations. Only very small residual diffraction-phase oscillations are visible, demonstrating the performance of OCT pulses. For $\SigmaGauss \in \{0.01,0.1,0.3\}\,\hbar k$, the populations and contrasts $(P_{\text{out}},C)$ are $(0.99998,0.99997)$, $(0.99309,0.99567)$, and $(0.87722,0.70157)$, respectively.}
\label{Fig:DifPhasen_3}
\end{figure}

\end{widetext}

\begin{widetext}

\begin{figure}[H]
\centering
% Column titles
\begin{minipage}[b]{0.05\textwidth}
\hfill
\end{minipage}
\begin{minipage}[b]{0.3\textwidth}
\centering
\textbf{$\SigmaGauss = 0.01 \hbar k$}
\end{minipage}
\begin{minipage}[b]{0.3\textwidth}
\centering
\textbf{$\SigmaGauss = 0.1 \hbar k$}
\end{minipage}
\begin{minipage}[b]{0.3\textwidth}
\centering
\textbf{$\SigmaGauss = 0.3 \hbar k$}
\end{minipage}

\vspace{0.2cm}

% First row title and figures
\begin{minipage}[b]{0.05\textwidth}
\centering
\raisebox{2cm}{\textbf{$\Phi_1$}} % Adjust the raisebox value to center vertically
\end{minipage}
\begin{minipage}[b]{0.3\textwidth}
\centering
\subfigure{\includegraphics[width=\textwidth]{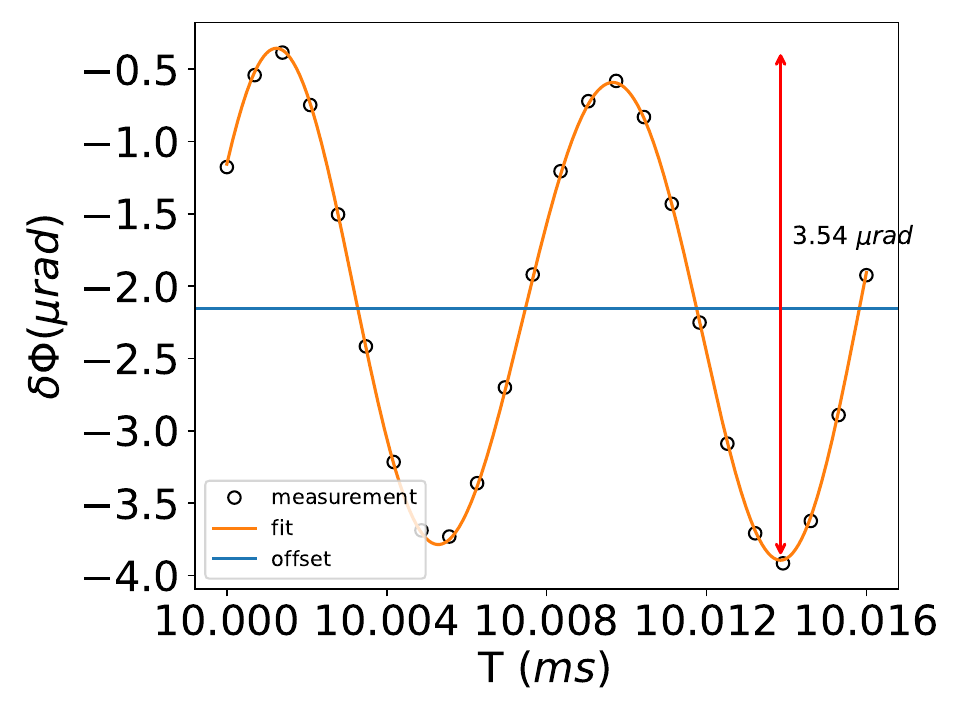}}
\end{minipage}
\begin{minipage}[b]{0.3\textwidth}
\centering
\subfigure{\includegraphics[width=\textwidth]{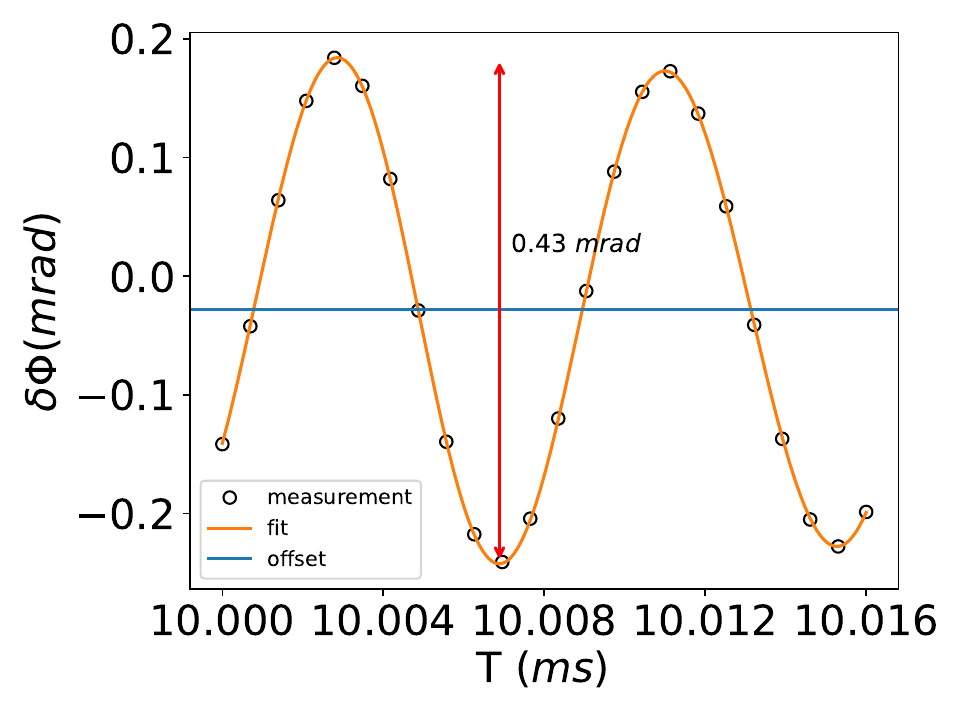}}
\end{minipage}
\begin{minipage}[b]{0.3\textwidth}
\centering
\subfigure{\includegraphics[width=\textwidth]{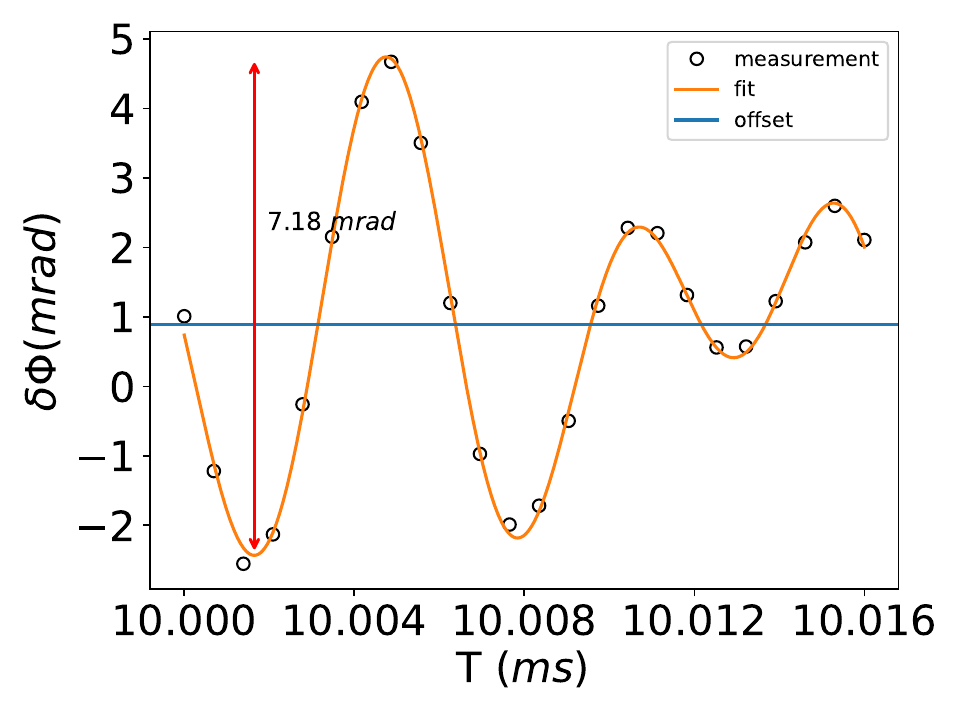}}
\end{minipage}

\vspace{0.2cm}

% Second row title and figures
\begin{minipage}[b]{0.05\textwidth}
\centering
\raisebox{2cm}{\textbf{$\Phi_2$}} % Adjust the raisebox value to center vertically
\end{minipage}
\begin{minipage}[b]{0.3\textwidth}
\centering
\subfigure{\includegraphics[width=\textwidth]{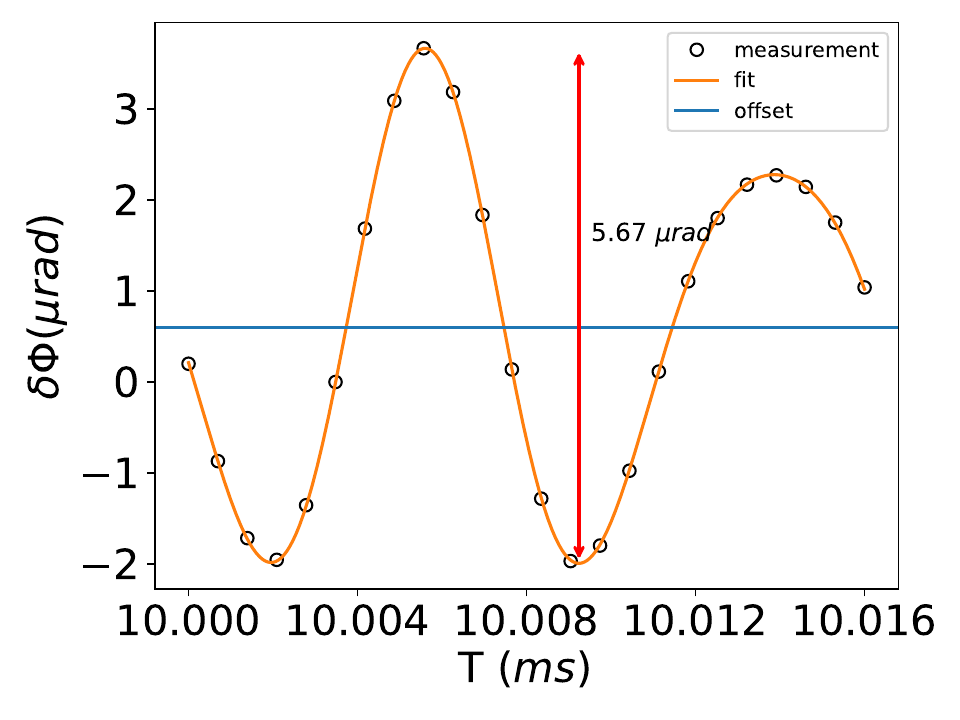}}
\end{minipage}
\begin{minipage}[b]{0.3\textwidth}
\centering
\subfigure{\includegraphics[width=\textwidth]{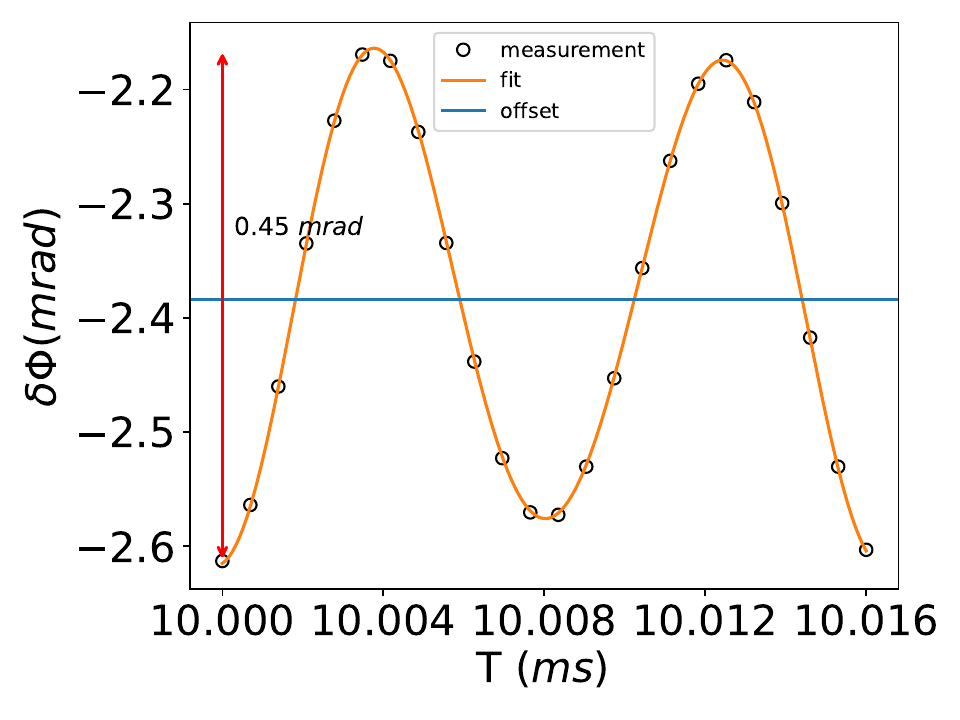}}
\end{minipage}
\begin{minipage}[b]{0.3\textwidth}
\centering
\subfigure{\includegraphics[width=\textwidth]{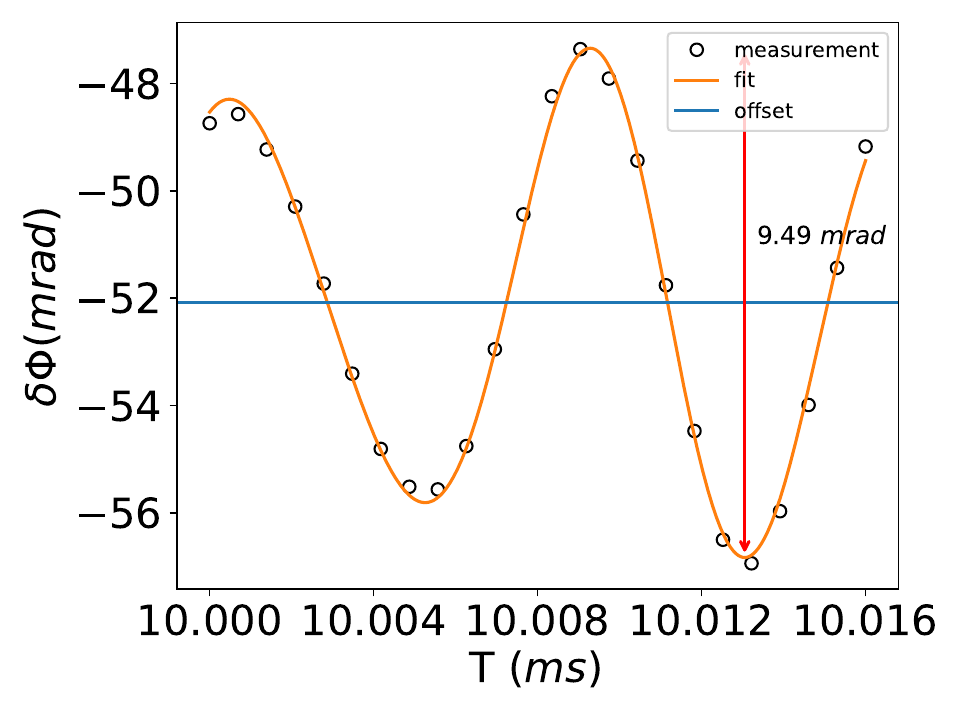}}
\end{minipage}
\caption{Diffraction phase for Bragg diffraction order $n=5$. The first and second rows show the diffraction phase at the first and second mid-fringe, respectively. The columns, from left to right, correspond to $\SigmaGauss \in \{0.01,0.1,0.3\}\,\hbar k$. In each plot, the solid blue line indicates the model shift, the empty circles represent the simulated data, and the orange curve corresponds to Eq.~\eqref{eq:OscFitModel} fitted to the simulations. Only very small residual diffraction-phase oscillations are visible, demonstrating the performance of OCT pulses. For $\SigmaGauss \in \{0.01,0.1,0.3\}\,\hbar k$, the populations and contrasts $(P_{\text{out}},C)$ are $(0.99994,0.99997)$, $(0.99157,0.98874)$, and $(0.78424,0.69266)$, respectively.}
\label{Fig:DifPhasen_5}
\end{figure}

\end{widetext}

We simulate the signal fringes for the MZ interferometer for Bragg orders $n=3,5$ by scanning 50 discrete values for the laser phase $\laserphase \in [0,2.4]$ rad via the second beam splitter in Fig.~\ref{Fig:MZI}. For each $\laserphase$, we then average over $N= 24\cdot 10^{6}$ samples picked from the momentum distribution of the incoming wave packet. This is necessary to ensure sub-mrad phase resolution of the interferometer fringes. After calibrating $P_a(\Phi)$ in Eq.~\eqref{eq:signalmodel} based on this data, we proceed to evaluate $\delta \Phi$ at the center of the fringe as sketched in Fig.~\ref{Fig:MidfringesSketch}, where the signal is maximally sensitive to changes in the phase. We determine the two first mid-fringe positions $\Phi_{\mathrm{L},1}$ and $\Phi_{\mathrm{L},2}$ with $P_a(\Phi_{\mathrm{L},i})= 0.5$ as highlighted in the figure.
Computing $\Phi^{\text{measured}}$ with the same resolution as the wave packet's momentum distribution as before yields the diffraction phase $\delta \Phi_i = \Phi^{\text{measured}}_i-\Phi_{\mathrm{L},i}$ at these points. This is what we refer to as \textit{measurement} in Figs.~\ref{Fig:DifPhasen_3} and \ref{Fig:DifPhasen_5}. 

Requiring $N= 24\cdot 10^{6}$ samples to achieve $\mu$rad phase resolution highlights the significant computational effort involved in the optimization. Therefore, it is more efficient to optimize individual operations within the interferometer rather than the entire sequence at a lower resolution for which we have obtained worse results. Attempting to optimize the full sequence with the same precision would necessitate a vastly greater number of samples, making it computationally infeasible. The optimization of the Beam splitter and Mirror, parallelizing the optimizations on a modern workstation CPUs took on the order of 4 days for each case, and the simulation 6 days.

%Expected error signal

To determine the phase accuracy of the OCT-enhanced Bragg interferometer, we determine the residual oscillation in the diffraction phase upon varying the time between the pulses, denoted as $T$ in Fig.~\ref{Fig:MZI}. This $T$-dependence of the interferometer signal due to the emergence of parasitic interferometers introduces potentially challenging systematic errors due to aliasing effects~\cite{ParkerMultiportBraggdiffraction, JanniPRL}. For the Bragg orders under consideration, the interference between the dominant parasitic paths and the main paths of the interferometer is described by oscillating terms due to the different accumulated phase with frequencies $8\omega_r(n+1)$ and $8\omega_r(n-1)$~\cite{Altin_AtomicGravimetryBragg2013, JanniPRL}
\begin{align}\label{eq:OscFitModel}
     \delta \Phi=& \xi_0 + \xi_1 \cos{((n-1) 8 \recoilfreq T + \nu_1)}\nonumber \\
     &+ \,\xi_2 \cos{((n+1) 8 \recoilfreq T + \nu_2)} .
\end{align}
Here, $\xi_i$ and $\nu_i$ are free fit parameters. We compute $\delta \Phi$ at the two mid-fringe positions for 24 values of $T\in [10,10.016]\, \mathrm{ms}$ to resolve one full oscillation period determined by $(n\pm1) 8 \recoilfreq $, and consider the specific case of $^{87}$Rb as an example.

%Results

Fig.~\ref{Fig:DifPhasen_3} shows the diffraction phase $\delta \Phi$ for Bragg order $n=3$, when scanning $T$ in the MZ interferometer. The top row displays the results for momentum widths $ \SigmaGauss \in \{0.01,0.1,0.3\} \hbar k$ for $\Phi_{\mathrm{L},1} $, while the bottom row is obtained with  $\Phi_{\mathrm{L},2} $. Scanning $T$, we observe residual oscillations in $\delta \Phi$. These oscillations are strongly suppressed by the use of OCT pulses, showing a peak-to-peak diffraction phase value for $\SigmaGauss = 0.01 \hbar k$ of a few $\mu$rad for the first fringe and below $\mu$rad for the second fringe. For $\SigmaGauss = 0.1\hbar k$, we have a peak-to-peak diffraction phase below mrad. The perfect fit of Eq.~\eqref{eq:OscFitModel} confirms that the diffraction phase oscillations originate from the residual couplings. For the case of $\SigmaGauss=0.3\hbar k$, we maintain a peak-to-peak diffraction phase of a few mrad, showing good performance even for relatively wide momentum distributions of the wavepacket. Nevertheless, for this case and at the working point $\Phi_{\mathrm{L},2} $, we have an almost $100$ mrad offset, which probably comes from a residual deformation of the fringe not captured by the signal model in Eq.~\eqref{eq:signalmodel}, making the fit of Eq.~\eqref{eq:OscFitModel} good but not perfect. In the caption, we also state the fraction of atoms that contribute to the interferometric signal, i.e., the ratio between the population measured in the output ports $P_{\text{out}}/P_{\text{total}}=N_a+N_b$ and the total number of atoms defined as $P_{\text{total}}\equiv 1$. This ratio tells us which percentage of atoms are measured in the main output ports with respect to the initial atoms. The combination of this ratio of atoms and the contrast $C$ shown in the caption, indicates that optimizing all pulses via OCT achieves a high contrast and low atom loss at the same time. This may be preferable to the deflection of atoms that populate parasitic paths from the interferometer, as suggested and done in Refs.~\cite{JanniPRL} and \cite{Pfeifferdichroic2025}.

Fig.~\ref{Fig:DifPhasen_5} is identical to Fig.~\ref{Fig:DifPhasen_3} but for $n=5$. Overall, we see the same behavior, with peak-to-peak values of a few $\mu$rad, below mrad, and a few mrad, respectively, for $ \SigmaGauss \in \{0.01,0.1,0.3\}\,\hbar k$. For $\SigmaGauss = 0.01 \hbar k$, we can compare with the results shown in Ref.~\cite{JanniPRL} for Gaussian pulses utilizing the so-called magic mirror. We obtain a better diffraction phase by a factor of around 8 and 100, respectively, for $\Phi_{\mathrm{L},2}$ and $\Phi_{\mathrm{L},1}$, with the advantage that we used a simpler estimator to fit the interferometer signal. For $\SigmaGauss = 0.3\hbar k$, the fitting error is smaller for $n=5$ compared to the case of $n=3$ and the fit offset is about $52$ mrad.

We showed that OCT pulses allow for $\mu$rad level of phase control if the atomic ensemble features velocity dispersions with $\SigmaGauss \leq 0.1 \hbar k$. This can set the requirements on the control of the velocity dispersion to ensure $\mu$rad-level phase control, not only for the peak-to-peak oscillations but also for the absolute shift. For $\SigmaGauss = 0.3\hbar k$, OCT pulses still suppress peak-to-peak oscillations to a few mrad, achieving good contrast, stability, and ratio of atoms that contribute to the interferometer signal. However, the fringe is still somewhat deformed compared to the two-port model, leading to the different offsets, especially for $\Phi_{\mathrm{L},2}$. The fact that the error of the fit to a two-level system is on the order of $\approx 10^{-7}$, $10^{-5}$ and $10^{-4}$ respectively for $ \SigmaGauss \in \{0.01,0.1,0.3\} \,\hbar k$  indicates that the phase control afforded by the OCT pulses extends over the entire fringe.

\section{Conclusions}\label{SectionConclusions}
In this work, we engineered  OCT-enhanced pulses of metrological relevance and demonstrated it by mitigating diffraction phase effects in high-order Bragg diffraction processes. So far, such pulses were promoted in the context of simply enhancing the contrast of atom interferometers. We showed the potential of ideally restoring a two-mode interferometry operation even for fundamentally multi-path, multi-port phenomena as the Bragg diffraction at high order. This brings this class of interferometers to the same level of their counterparts in the Raman regime, so far considered as more immune to transitions to  unwanted states.

As an illustration, we have benchmarked the performance of third- and fifth-order Bragg beam splitters and mirrors using OCT pulses against optimal Gaussian pulses. We found that while the fidelity of Gaussian pulses suffers significantly due to coupling to unwanted states, as well as from finite velocity widths of the atomic ensemble, OCT pulses ensure excellent diffraction efficiencies.

We found that OCT pulses generally suppress diffraction phases below the $\mu$rad level for sufficiently cold clouds, achieving a performance of a few $\mu$rad or below for $\SigmaGauss = 0.01\hbar k$ and growing only to a few mrad if we account for momentum widths up to $\SigmaGauss = 0.3 \hbar k$. These diffraction phases have been calculated for the midfringe of the signal of the interferometer, nevertheless, the phase-scan is close to a two-mode signal for any value of the imprinted phase, extending the phase control over the entire fringe. In a next step, one can now assess the robustness against imperfections in the control parameters, similar to Ref. \cite{Li_2026} where comparable techniques are used. 

Our work contributes to the pursuit of atom interferometers with $\mu$rad precision, which is a requirement for challenging applications in finding physics beyond the standard model~\cite{Safronova2018}  or detecting gravitational waves~\cite{Abdalla2025}.  

\vspace*{0.5cm}

\begin{acknowledgments}
We want to thank Stuart Szigeti, Andre Carvalho, Jack Saywell and Richard Taylor for early discussions about the pulse optimizations. This work was funded by the Deutsche Forschungsgemeinschaft (German Research Foundation) under Germany’s Excellence Strategy (EXC-2123 QuantumFrontiers Grants No. 390837967) and the German Space Agency (DLR) with funds provided by the German Federal Ministry for Economic Affairs and Climate Action (BMWK) (German Federal Ministry of Education and Research (BMBF)) due to an enactment of the German Bundestag under Grant No. 50WM2253A (AI-Quadrat) and Project-ID 274200144– SFB 1227 (DQ-mat, project A05). NG acknowledges funding by the AGAPES project - grant No. 530096754 within the ANR-DFG 2023 Programme. J.-N. K.-S. and N.G. acknowledge support from QuantumFrontiers through the QuantumFrontiers Entrepreneur Excellence Programme (QuEEP). J.-N. K.-S. and N.G. acknowledge funding from the EU project CARIOQA-PMP (No. 101081775).

\end{acknowledgments}

\begin{section}*{Conflict of interest}

The authors have no conflicts to disclose

\end{section}

\begin{section}*{Data availability}

The data that supports the findings of this study are available from the corresponding authors upon reasonable request

\end{section}

\appendix

\section{Relative phases induced by momentum detuning}
\label{App:DynPhases}

In this work, we apply OCT on the level of the unitary matrix, i.e., optimizing the fidelity, as this is more complete than considering only populations. For the OCT pulses, where, due to the freedom over both the Rabi frequency and $\smalldet(t)$, optimizing only the populations results in a random momentum-dependent phase imprint that destroys the interferometer signal. Therefore, for OCT pulses, it is highly preferable to optimize the unitary matrix. For the Gaussian pulses, due to their enforced symmetry and because $\smalldet(t) = 0$, optimizing populations creates an output state with a great splitting ratio that is otherwise very sensitive to the Doppler detuning and can be compensated if the second beam splitter is identical to the first. This phase and the extent of its effects on atom interferometry have previously been studied for Raman pulses with box-shaped time profiles~\cite{DosSantosLimitsSymmetry,morel2020velocitydependentphase}.

\begin{figure}[H]

\begin{minipage}[b]{0.4\textwidth}
\centering
\subfigure{\includegraphics[width=\textwidth]{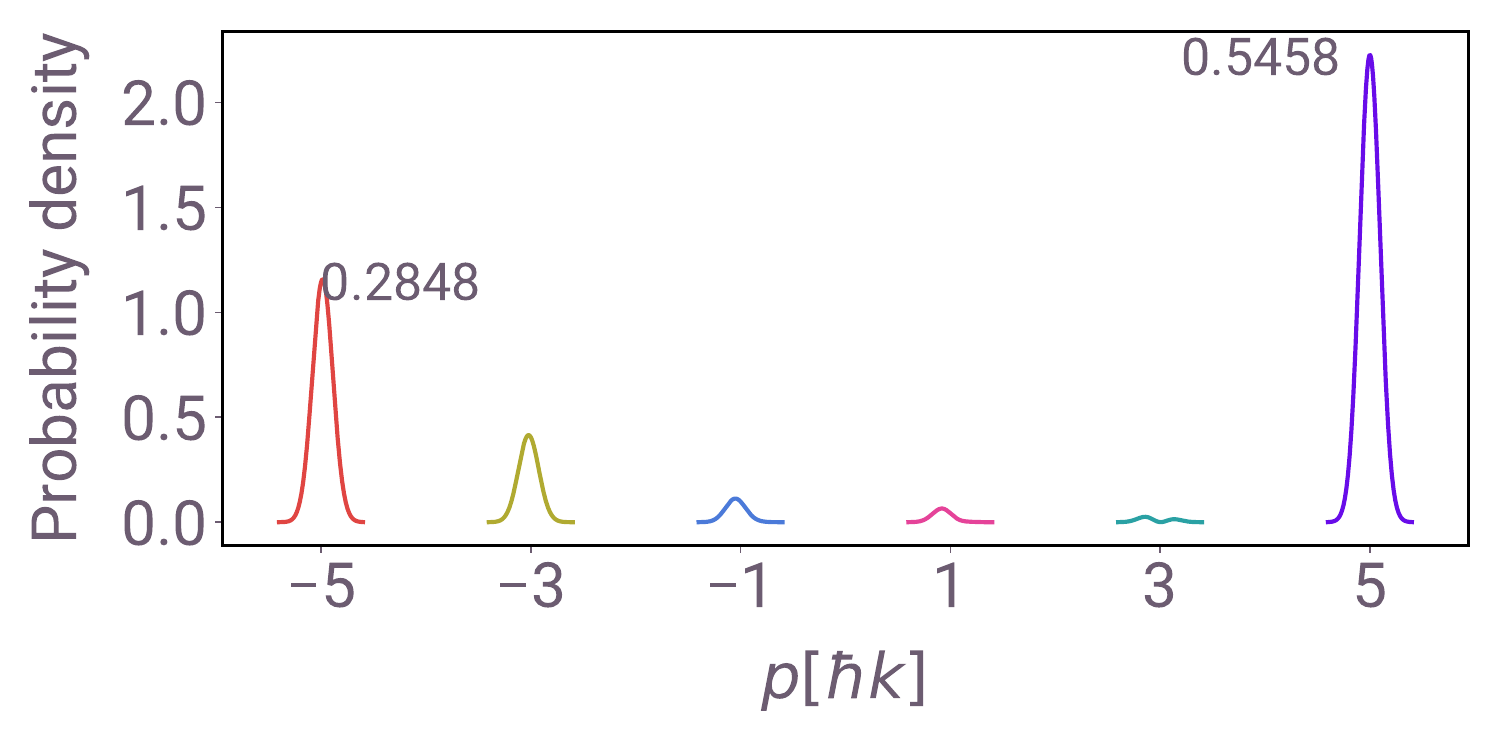}}
\end{minipage}
\vspace{0.1cm}

% Second row title and figures
\begin{minipage}[b]{0.4\textwidth}
\centering
\subfigure{\includegraphics[width=\textwidth]{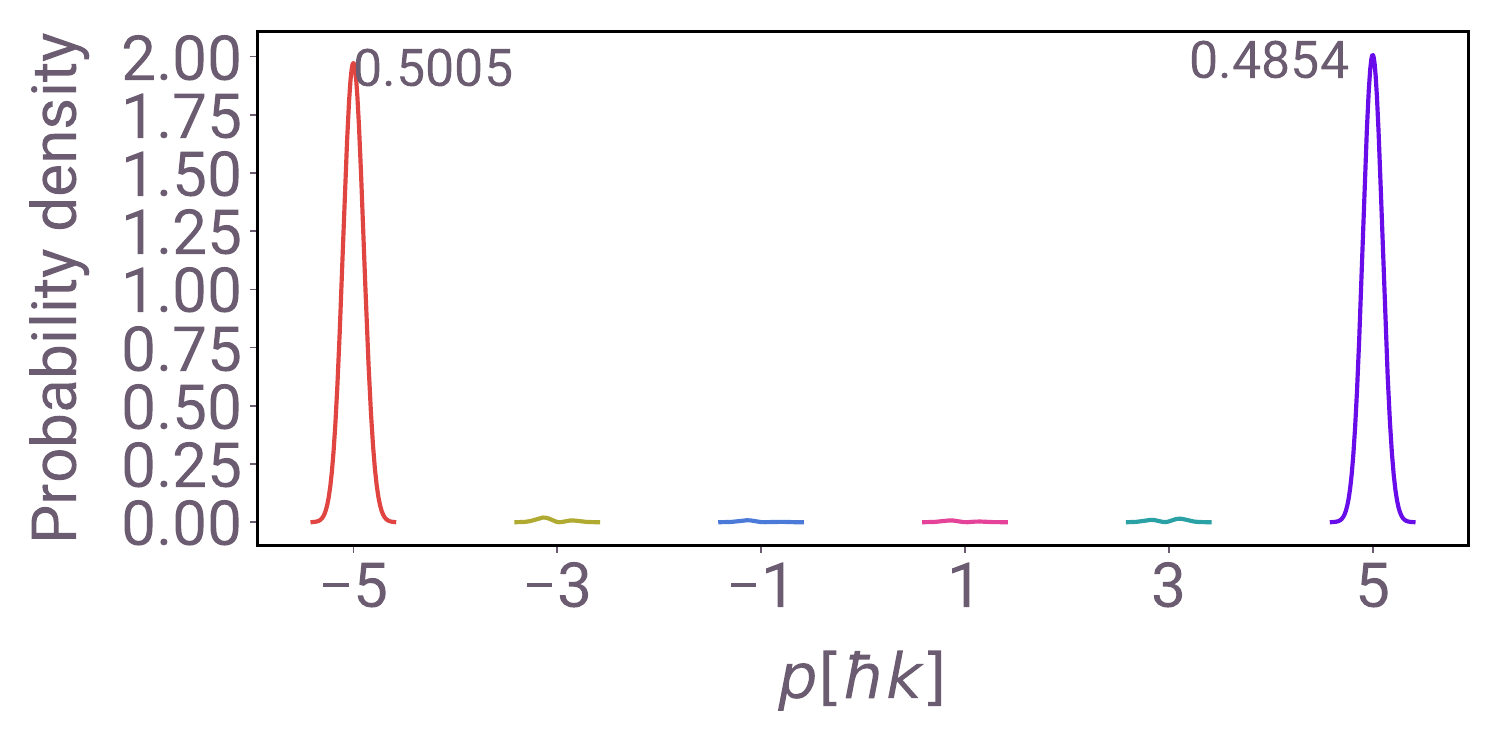}}
\end{minipage}
\caption{Momentum distribution of an incoming state $\ket{-5 \hbar k}$ with momentum width $\SigmaGauss = 0.1 \hbar k$ after interacting with the laser that generates a beam splitter pulse. \textbf{Top}: Gaussian pulse optimized for the unitary submatrix of the interaction, yielding $\left(\Omega_0, \tau\right)= \left(34\,\recoilfreq,0.145\,\recoilfreq^{-1}\right)$. \textbf{Bottom}: Gaussian pulse optimized for an equal population split between the states $\ket{5 \hbar k}$ and $\ket{-5 \hbar k}$, we obtain $\left(\Omega_0, \tau\right)= \left(30.952\,\recoilfreq,0.210 \,\recoilfreq^{-1}\right)$.}\label{Fig:PopComparison}
\end{figure}

As an example, we will study the population for the case of $\SigmaGauss= 0.1 \hbar k$ and $n = 5$. In Fig.~\ref{Fig:PopComparison}, we show the population transfer of a beam splitter when optimizing the population transfer (Bottom), which for a mirror corresponds to the population in the target state and for a beam splitter is described by
\begin{align} \label{eq:PBScost}
\text{Pcost}_{BS} =& \frac{1}{N} \sum_{i=1}^N \left|0.5 - \|\bra{n\hbar k} \hat{U}^{(i)}\ket{-n \hbar k}\|\right| \nonumber\\
&+ \left|0.5 - \|\bra{-n\hbar k} \hat{U}^{(i)}\ket{-n \hbar k}\|\right| 
\end{align}
and when using the cost function in Eq.~\eqref{eq:avgInfidelity} (Top). One can see an improvement in population transfer when optimizing the populations; however, this comes at the cost of a worse individual beam splitter in terms of fidelity as shown in Tab.~\ref{Tab:GP}.

The fidelities of the Gaussian pulses optimizing populations ($G_P$) against the fidelities of the Gaussian pulses optimizing the fidelity ($G_U$) are shown in Tab.~\ref{Tab:GP}. We first note that the performance of the OCT pulses is significantly higher in all the metrics, as seen in Tab~\ref{Tab:n3n5}. Nevertheless, we also seek to understand the differences between optimizing the population or the unitary matrix for the Gaussian pulses and compare both approaches to the OCT pulses, both for completeness and to gain insight into the underlying physics.

For all cases evaluated in Tab.~\ref{Tab:GP}, the beam splitter fidelity is higher or equal when the fidelity itself is optimized, as it was expected. For the mirror, however, the weight of the parasitic paths was taken differently for both optimizations, so a direct comparison should not be made. Nevertheless, they have similar performance. 

Since the behavior very similar considering different combinations of $(\SigmaGauss, n)$, we focus on the example $\SigmaGauss= 0.1 \hbar k$ and $n = 5$. In this case, the fidelity of the beam splitter, the mirror and the combined beam splitter-mirror sequence are all higher when the fidelity of the individual operations was optimized. However, the fidelity of the full interferometer operation is higher when the populations are optimized. 

From this, one might conclude that, in general, the performance of the Gaussian pulses for the full interferometer -considering the effective unitary- is better when the populations are optimized. However, we recall to the reader that we do not include asymmetries like intensity fluctuations, expansion of the cloud between pulses, etc. (cf. Refs. \cite{DosSantosLimitsSymmetry} and \cite{morel2020velocitydependentphase}). Yet, the compensation in Tab.~\ref{Tab:GP} for the full interferometers relies on the symmetry of the system which will not always hold in realistic scenarios~\cite{QCTRLAI,DosSantosLimitsSymmetry}. 

\newcommand{\getColumnValuess}[2]{% #1 = row number, #2 = column name
    \csvreader[head to column names]{TESTNewtabledata5U.csv}{}{%
        \ifnum\thecsvrow=#1 % Check if it's the x-th row
            \pgfmathparse{#2} % Parse the value of the specified column
            \def\ColumnValue{\pgfmathresult} % Store the value in the macro
        \fi
    }%
}
\newcommand{\getColumnValuesss}[2]{% #1 = row number, #2 = column name
    \csvreader[head to column names]{TESTNewtabledata3U.csv}{}{%
        \ifnum\thecsvrow=#1 % Check if it's the x-th row
            \pgfmathparse{#2} % Parse the value of the specified column
            \def\ColumnValue{\pgfmathresult} % Store the value in the macro
        \fi
    }%
}

\begin{table}[h!]
\centering
\begin{tabular}{c c >{\columncolor{gray!15}}c >{\columncolor{blue!8}}c >{\columncolor{gray!15}}c >{\columncolor{blue!8}}c}
%\hline
&&\multicolumn{2}{c}{$n=3$}&\multicolumn{2}{c}{$n=5$}\\
\multirow{1}{*}{}  & $\SigmaGauss\;[\hbar k]$ & G$_U$ & G$_\text{P}$ & G$_U$ & G$_\text{P}$ \\
\midrule
& $0.01$ & \getColumnValues{4}{\BS}\ColumnValue& \getColumnValuesss{4}{\BS }\ColumnValue & \getColumnValue{4}{\BS}\ColumnValue & \getColumnValuess{4}{\BS }\ColumnValue \\
$\hat{U}_\mathrm{BS}$ & $0.1$ & \getColumnValues{5}{\BS}\ColumnValue & \getColumnValuesss{5}{\BS} \ColumnValue & \getColumnValue{5}{\BS}\ColumnValue & \getColumnValuess{5}{\BS}\ColumnValue \\
& $0.3$ & \getColumnValues{6}{\BS}\ColumnValue & \getColumnValuesss{6}{\BS}\ColumnValue & \getColumnValue{6}{\BS}\ColumnValue & \getColumnValuess{6}{\BS}\ColumnValue \\  
\midrule
& $0.01$ & \getColumnValues{4}{\M}\ColumnValue& \getColumnValuesss{4}{\M }\ColumnValue & \getColumnValue{4}{\M}\ColumnValue & \getColumnValuess{4}{\M }\ColumnValue \\
$\hat{U}_\mathrm{M}$ & $0.1$ & \getColumnValues{5}{\M}\ColumnValue & \getColumnValuesss{5}{\M} \ColumnValue & \getColumnValue{5}{\M}\ColumnValue & \getColumnValuess{5}{\M}\ColumnValue \\
& $0.3$ & \getColumnValues{6}{\M}\ColumnValue & \getColumnValuesss{6}{\M}\ColumnValue & \getColumnValue{6}{\M}\ColumnValue & \getColumnValuess{6}{\M}\ColumnValue \\  
\midrule
& $0.01$ & \getColumnValues{4}{\MBS}\ColumnValue& \getColumnValuesss{4}{\MBS }\ColumnValue & \getColumnValue{4}{\MBS}\ColumnValue & \getColumnValuess{4}{\MBS }\ColumnValue \\
$\hat{U}_\mathrm{M}\hat{U}_\mathrm{BS}$ & $0.1$ & \getColumnValues{5}{\MBS}\ColumnValue & \getColumnValuesss{5}{\MBS} \ColumnValue & \getColumnValue{5}{\MBS}\ColumnValue & \getColumnValuess{5}{\MBS}\ColumnValue \\
& $0.3$ & \getColumnValues{6}{\MBS}\ColumnValue & \getColumnValuesss{6}{\MBS}\ColumnValue & \getColumnValue{6}{\MBS}\ColumnValue & \getColumnValuess{6}{\MBS}\ColumnValue \\  
\midrule
& $0.01$ & \getColumnValues{4}{\BSMBS}\ColumnValue& \getColumnValuesss{4}{\BSMBS }\ColumnValue & \getColumnValue{4}{\BSMBS}\ColumnValue & \getColumnValuess{4}{\BSMBS }\ColumnValue \\
$\hat{U}_\mathrm{BS}\hat{U}_\mathrm{M}\hat{U}_\mathrm{BS}$ & $0.1$ & \getColumnValues{5}{\BSMBS}\ColumnValue & \getColumnValuesss{5}{\BSMBS} \ColumnValue & \getColumnValue{5}{\BSMBS}\ColumnValue & \getColumnValuess{5}{\BSMBS}\ColumnValue \\
& $0.3$ & \getColumnValues{6}{\BSMBS}\ColumnValue & \getColumnValuesss{6}{\BSMBS}\ColumnValue & \getColumnValue{6}{\BSMBS}\ColumnValue & \getColumnValuess{6}{\BSMBS}\ColumnValue \\  
\bottomrule
\end{tabular}
\caption{Comparison of fidelities, see Eq.~\eqref {eq:Fidelity}, of a beam splitter, a mirror, their combination, and the full interferometer for Gaussian pulses of Bragg orders $n=3,5$ obtained with two different cost functions. Columns $G_U$ are obtained using Eq.~\eqref{eq:avgInfidelity} and are identical to Tab.~\ref{Tab:n3n5}. Columns $G_\text{P}$ represent fidelities obtained when optimizing for population transfer, which correspond to the population of the target state for mirrors and for beam splitters is described in Eq.~\eqref{eq:PBScost}. The initial state is the same as in Tab.~\ref{Tab:n3n5} in the main text. The background colors (gray for $G_U$ and blue for $G_\text{P}$) are given to facilitate the reading.  }
\label{Tab:GP}
\end{table}

Finally, it may appear counterintuitive that the optimizer, when minimizing the cost in Eq.~\eqref{eq:avgInfidelity}, selects a solution that does not also maximize the population in the main interferometer arms. After all, the ideal unitary of a beam splitter divides the population of the ground and excited states equally. 

In case of OCT pulses the optimization algorithm has sufficient degrees of freedom to maximize both the population transfer and making the state insensitive to the momentum-dependent phase. Yet, this is not possible for a simple Gaussian pulse as defined in Eq.~\eqref{eq:GaussianPulse}. Therefore, as hinted by Tab.~\ref{Tab:GP}, which Gaussian pulse has a better performance will highly depend on the symmetry of the interferometer~\cite{DosSantosLimitsSymmetry}.

\section{Cost function for Gaussian case Mirror }
\label{App:CostGaussianMirror}

As motivated in the main text, and further confirmed in appendix~\ref{App:DynPhases}, the population in the parasitic ports after the first beam splitter is substantial for the optimized Gaussian pulses. Therefore, this had to be taken into account in the optimization of the mirror pulse. To do so, we weighted the population of the parasitic ports in each optimization as follows:

\begin{align}\label{eq:costMirrorGaussian}
    \text{cost}\left(\hat{U}_M\right) =&\frac{1}{N} \sum_{i=1}^N 1 -\frac{1}{2}\| \mathrm{Tr}\left(\hat{U}_M^{\dagger} \hat{U}^{(i)}\right) \| \nonumber \\
    &+ W\frac{1}{2}\| \mathrm{Tr}\left(\hat{\mathcal{U}}_{M}^{\dagger} \hat{\mathcal{U}}^{(i)}\right) \|,
\end{align}
where, as in the main text, $\hat{U}_M$ denotes the target unitary mirror for the main ports. $\hat{\mathcal{U}}_M$ in the second term of the sum in Eq.~\eqref{eq:costMirrorGaussian} represents the ideal mirror interaction for the main parasitic paths. This contribution to the cost function is necessary because the presence of atoms in these paths is non-negligible. Accordingly, the cost function penalizes the reflection of those atoms into the output ports (see Fig.~\ref{Fig:MZI}). The weight $W$, which determines how strongly the populations in the parasitic paths contribute to the cost function, is chosen to correspond to the cost associated with the optimization of the corresponding beam splitter for the same $\SigmaGauss$. Thus, the better the first beam splitter performs, the smaller W is. The values of W can be extracted from Tab.~\ref{Tab:n3n5} as the cost is the distance from 1 of the fidelity.

%---------------------------
\section{Comparison of population transfer for different cases }
%---------------------------
\label{App:populationcomparison}

In Sec.~\ref{PulseFidelity}, we compared the population transfer of an incoming state with momentum width $\SigmaGauss = 0.1\hbar k$ and Bragg order $n=5$, using an optimized Gaussian pulse and an OCT pulse. For completeness, we present in this appendix the population transfer for all pairs of $ \SigmaGauss \in \{0.01,0.1,0.3\}\, \hbar k$ and Bragg orders $n= 3,5$ examined in this article, for both optimized Gaussian pulses and OCT pulses, and for both mirrors and beam splitters.

In Fig.~\ref{Fig:Mirrorn3}(\ref{Fig:Mirrorn5}), we show the momentum distribution of an incoming state $\ket{-3 \hbar k}$($\ket{-5 \hbar k}$) after interacting with the laser that generates a mirror. The first row corresponds to the optimized Gaussian pulse, and the second row corresponds to the OCT pulse. Each column shows a different value of the momentum width $\SigmaGauss$ of the incoming state. We observe that the OCT pulses achieve a better population transfer: for $\SigmaGauss = 0.01 \hbar k$, the difference is small, but it exceeds $20 \%$($25 \%$) for $\SigmaGauss = 0.3 \hbar k$.

In Fig.~\ref{Fig:BSn3}(\ref{Fig:BSn5}), we illustrate the momentum distribution of an incoming state $\ket{-3 \hbar k}$($\ket{-5 \hbar k}$) after interacting with the laser that generates a beam splitter. The first row corresponds to the optimized Gaussian pulse, and the second row corresponds to the OCT pulse. Each column shows a different value of the momentum width distribution of the incoming state $\SigmaGauss$. We observe how the OCT pulses are closer to achieving an equal population split between the main ports. The momentum distribution for OCT pulses is approximately Gaussian in $p$, for $ \SigmaGauss \in \{0.01,0.1\}\, \hbar k$, but deviates from that shape for $\SigmaGauss = 0.3 \hbar k$. Moreover, it shows a considerable population in the parasitic ports. In the Gaussian case, for $\SigmaGauss = 0.3 \hbar k$ it is also noteworthy that the Gaussian pulse optimizer tends to reproduce on average the behavior of a beam splitter by effectively creating a 'bad' mirror, as can be seen from the momentum selectivity of the pulse.

\begin{widetext}

\begin{figure}[H]
\centering
\begin{minipage}[c]{0.05\textwidth} % narrow column for label
    \centering
    \rotatebox{90}{3rd order Bragg mirror}
\end{minipage}%
\begin{minipage}[c]{0.95\textwidth} % the rest of the figure
    % --- Your current figure content goes here ---
    \begin{minipage}[b]{0.32\textwidth}
        \centering
        \textbf{$\SigmaGauss = 0.01 \hbar k$}
    \end{minipage}
    \begin{minipage}[b]{0.32\textwidth}
        \centering
        \textbf{$\SigmaGauss = 0.1 \hbar k$}
    \end{minipage}
    \begin{minipage}[b]{0.32\textwidth}
        \centering
        \textbf{$\SigmaGauss = 0.3 \hbar k$}
    \end{minipage}

    \vspace{0.2cm}

    % First row
    \begin{minipage}[b]{0.32\textwidth}
        \centering
        \includegraphics[width=\textwidth]{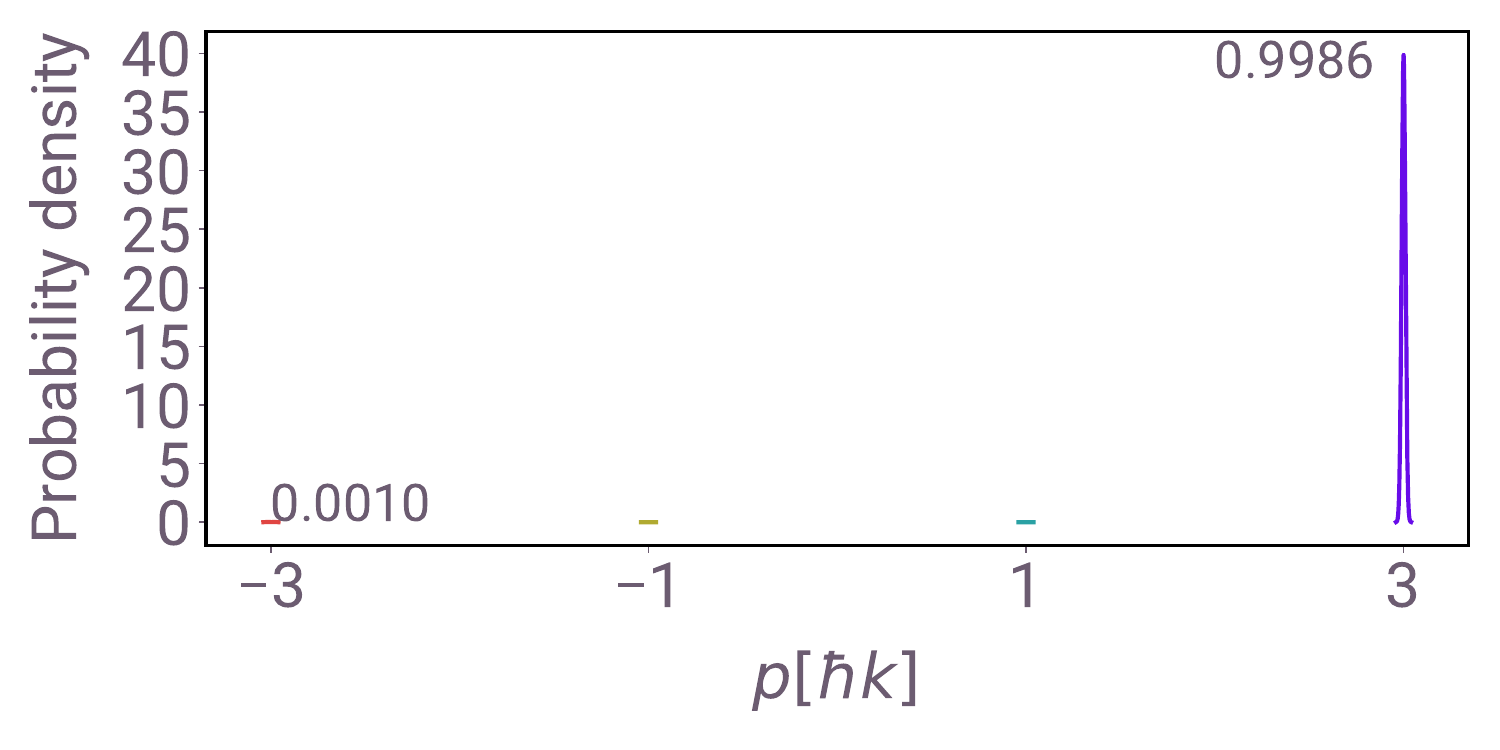}
    \end{minipage}
    \begin{minipage}[b]{0.32\textwidth}
        \centering
        \includegraphics[width=\textwidth]{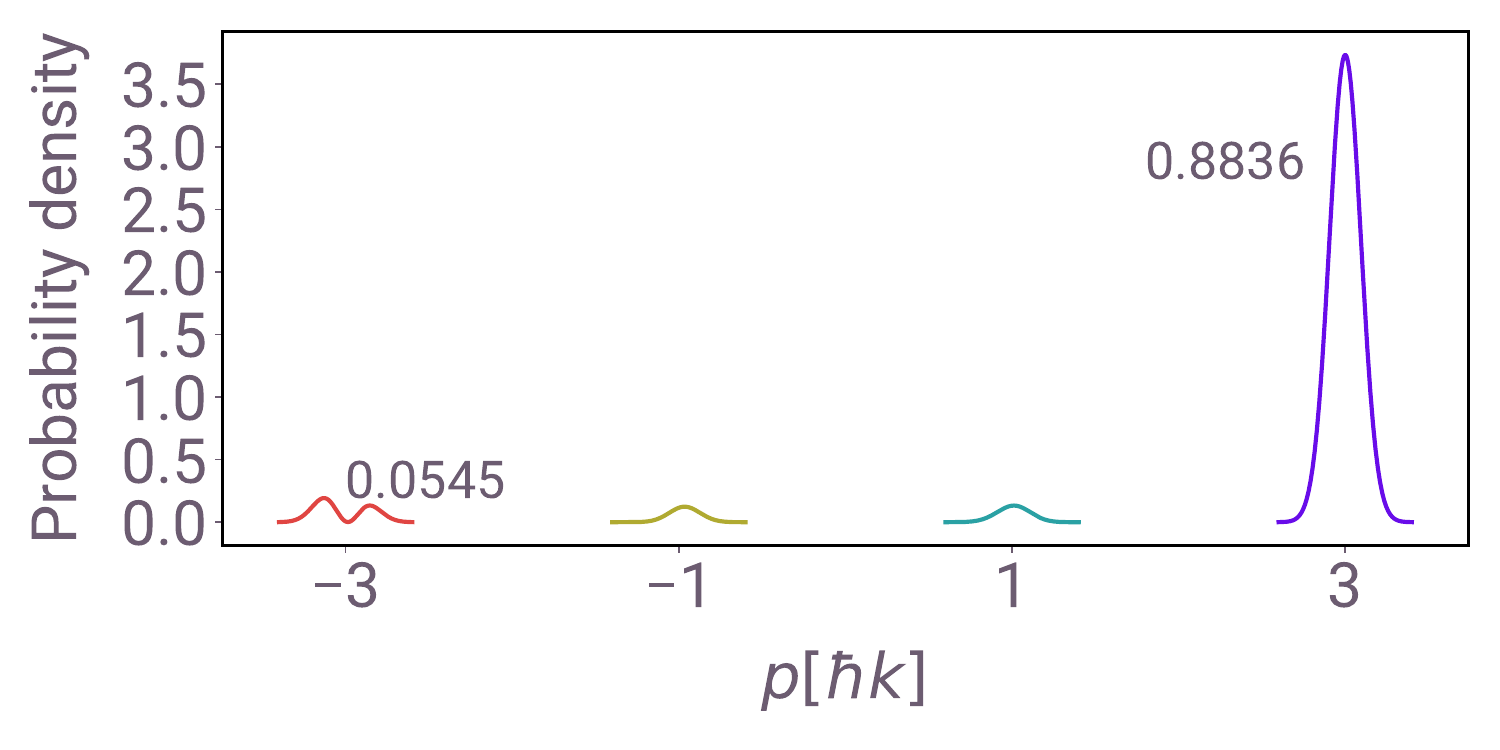}
    \end{minipage}
    \begin{minipage}[b]{0.32\textwidth}
        \centering
        \includegraphics[width=\textwidth]{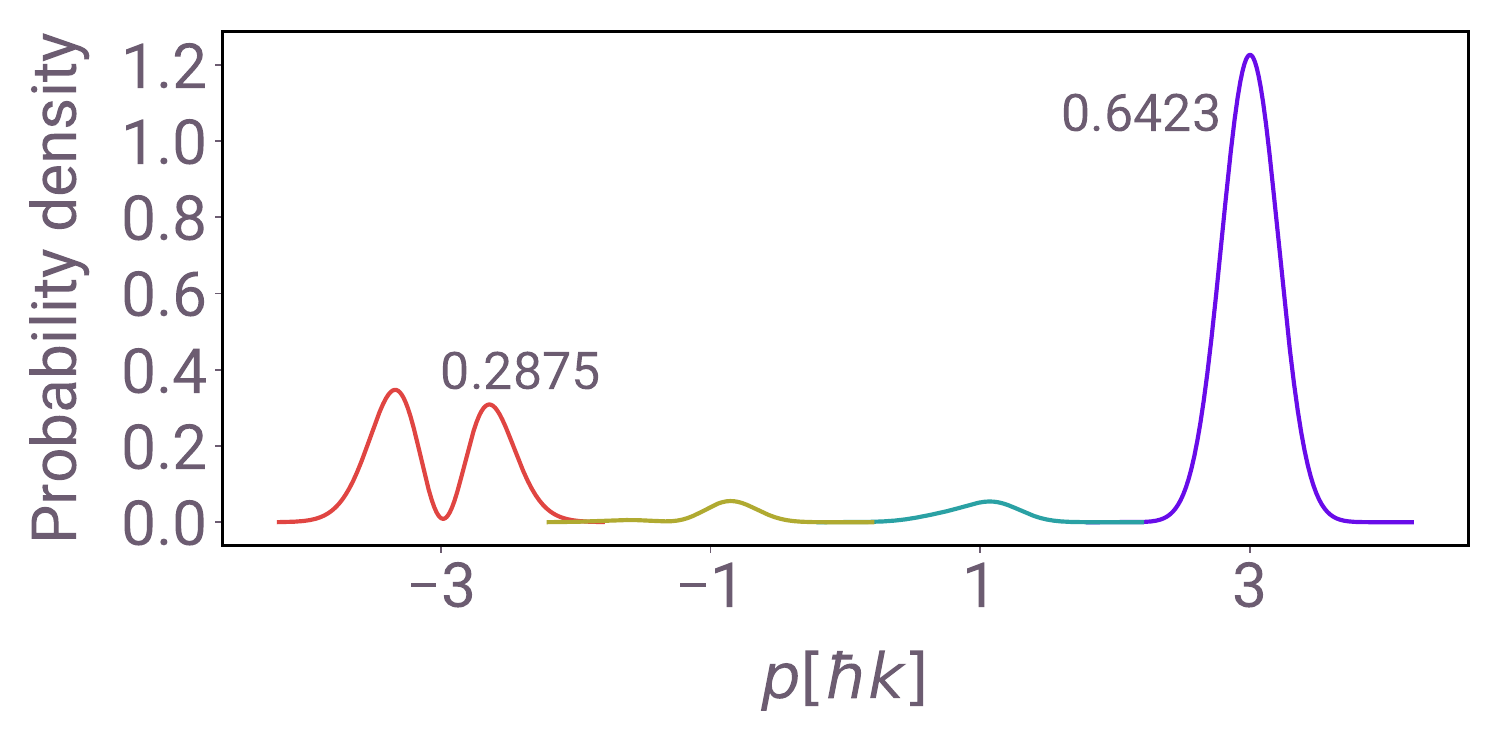}
    \end{minipage}

    \vspace{0.2cm}

    % Second row
    \begin{minipage}[b]{0.32\textwidth}
        \centering
        \includegraphics[width=\textwidth]{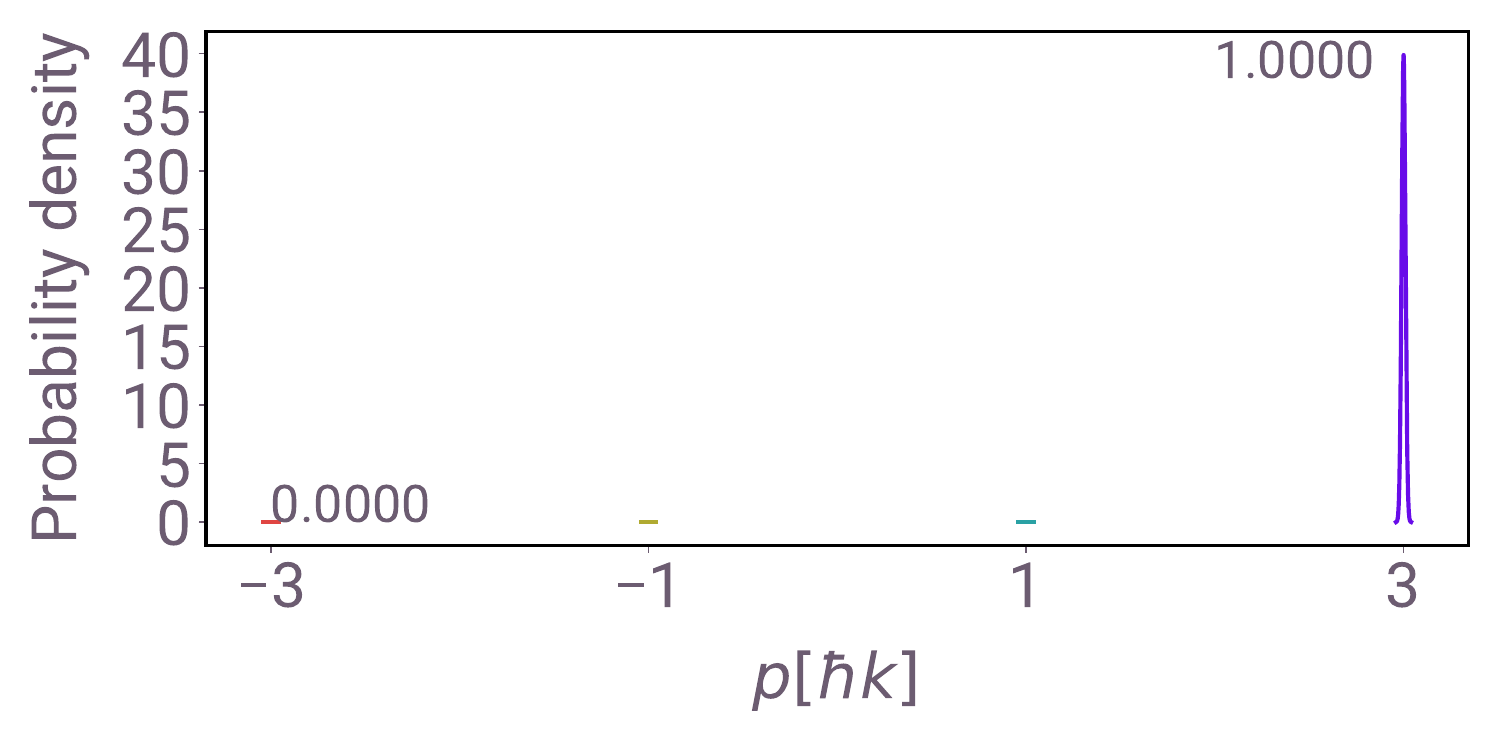}
    \end{minipage}
    \begin{minipage}[b]{0.32\textwidth}
        \centering
        \includegraphics[width=\textwidth]{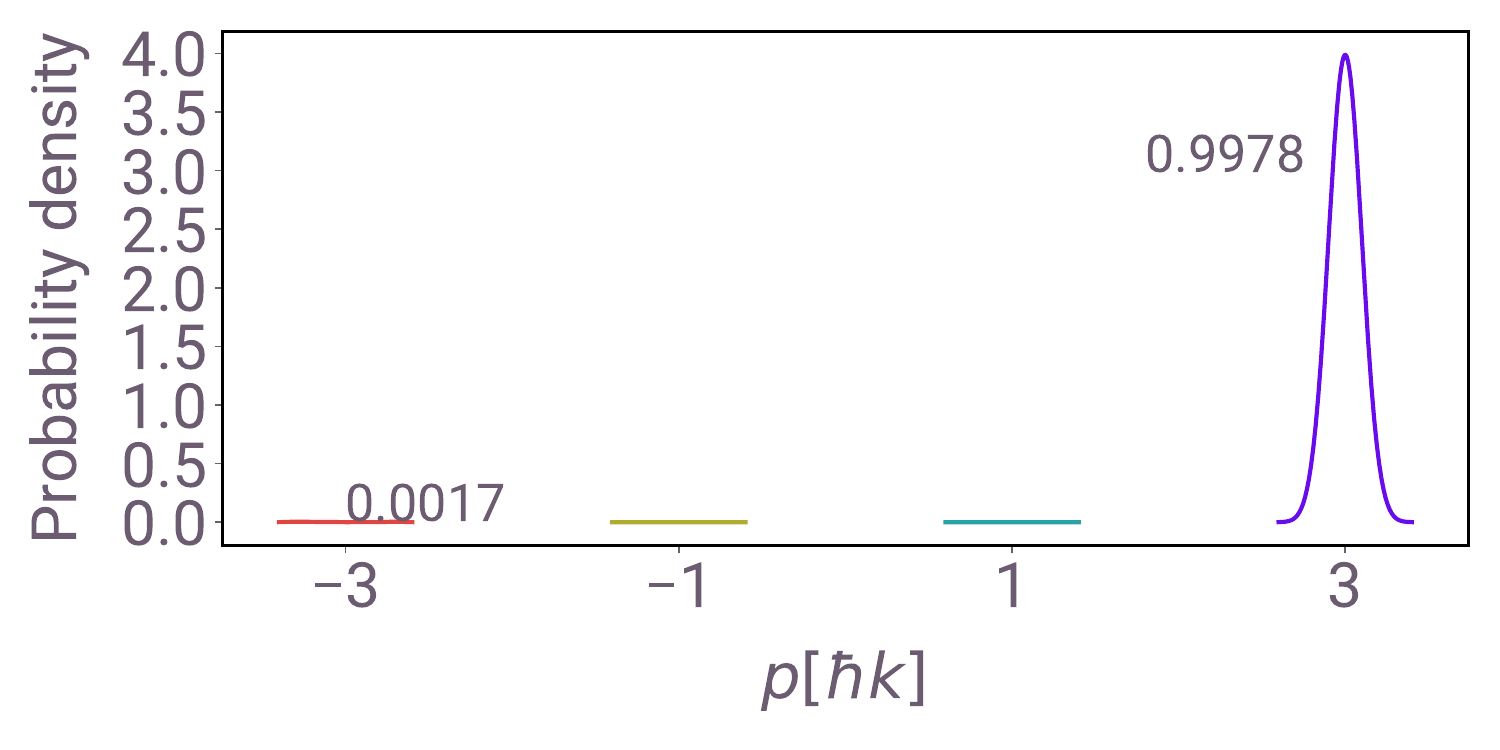}
    \end{minipage}
    \begin{minipage}[b]{0.32\textwidth}
        \centering
        \includegraphics[width=\textwidth]{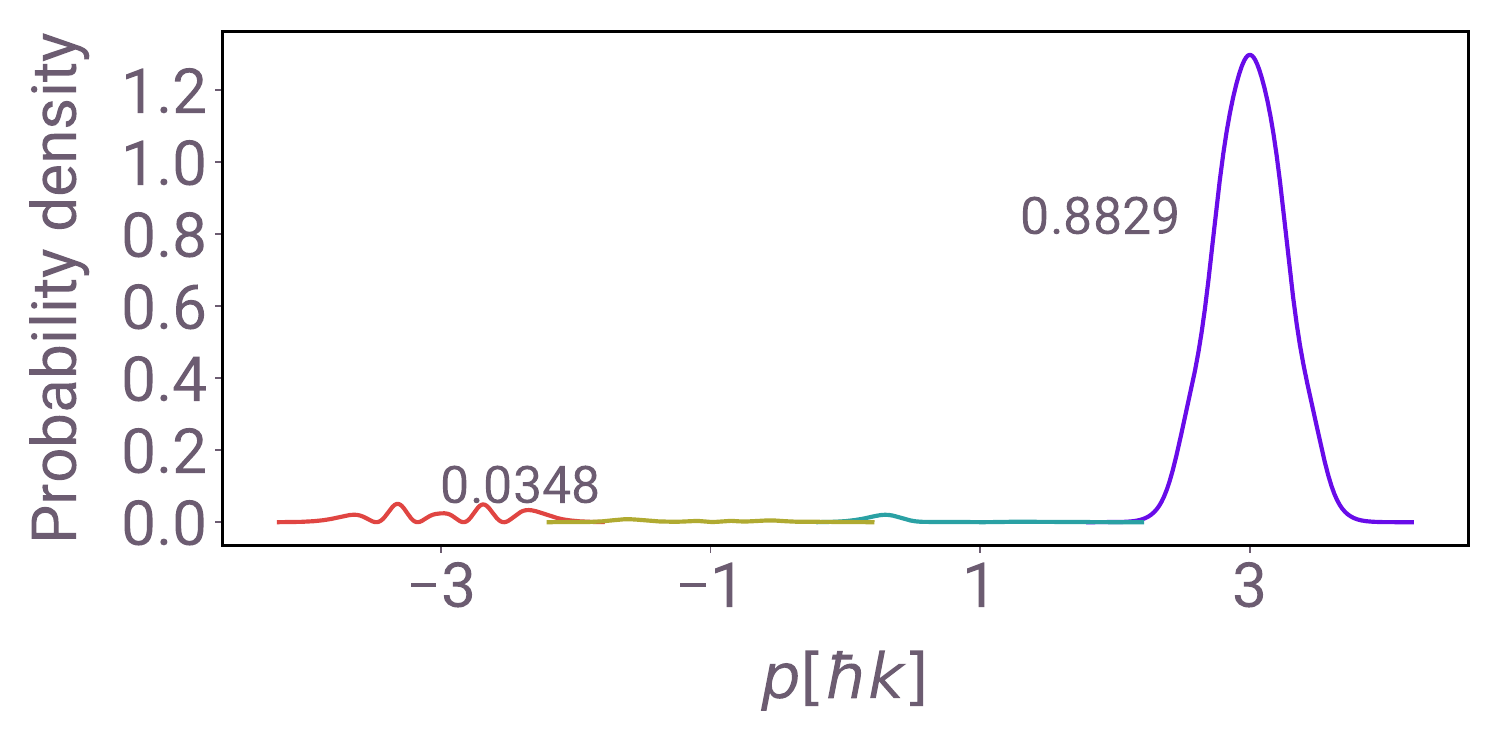}
    \end{minipage}
\end{minipage}

\caption{Momentum distribution of an incoming state $\ket{-3 \hbar k}$ after 3rd order Bragg mirror pulse. A Gaussian optimized pulse (top row) is compared to an OCT pulse (bottom row). The columns correspond to $ \SigmaGauss \in \{0.01,0.1,0.3\} \hbar k$. The population values of the main ports are indicated in the figures. The Gaussian pulse parameters are, in order, for the first row $\left(\Omega_0, \tau\right)=\{\left(12.81,0.463\right), \left(14.52,0.349\right),\left(15.09,0.337\right)\}$, where $\Omega_0$ is given in units of $\recoilfreq$ and $\tau$ in units of $\recoilfreq^{-1}$.}

% \caption{Momentum distribution of an incoming state $\ket{-3 \hbar k}$ after interacting with the laser that generates a mirror pulse. The first row shows the result for a Gaussian-optimized pulse and the second row shows the result for an OCT pulse. The columns correspond to $ \SigmaGauss \in \{0.01,0.1,0.3\} \hbar k$. The population values of the main ports are indicated in the figures. The Gaussian pulse parameters are, in order, for the first row $\left(\Omega_0, \tau\right)=\{\left(12.81,0.463\right), \left(14.52,0.349\right),\left(15.09,0.337\right)\}$, where $\Omega_0$ is given in units of $\recoilfreq$ and $\tau$ in units of $\recoilfreq^{-1}$.}
\label{Fig:Mirrorn3}
\end{figure}

\begin{figure}[H]
\centering
\begin{minipage}[c]{0.05\textwidth}
\rotatebox{90}{5th order Bragg mirror}
\end{minipage}%
\begin{minipage}[c]{0.93\textwidth}
\centering
% Column titles
\begin{minipage}[b]{0.32\textwidth}
\centering
\textbf{$\SigmaGauss = 0.01 \hbar k$}
\end{minipage}
\begin{minipage}[b]{0.32\textwidth}
\centering
\textbf{$\SigmaGauss = 0.1 \hbar k$}
\end{minipage}
\begin{minipage}[b]{0.32\textwidth}
\centering
\textbf{$\SigmaGauss = 0.3 \hbar k$}
\end{minipage}

\vspace{0.2cm}

% First row: Gaussian-optimized pulse
\begin{minipage}[b]{0.32\textwidth}
\centering
\includegraphics[width=\textwidth]{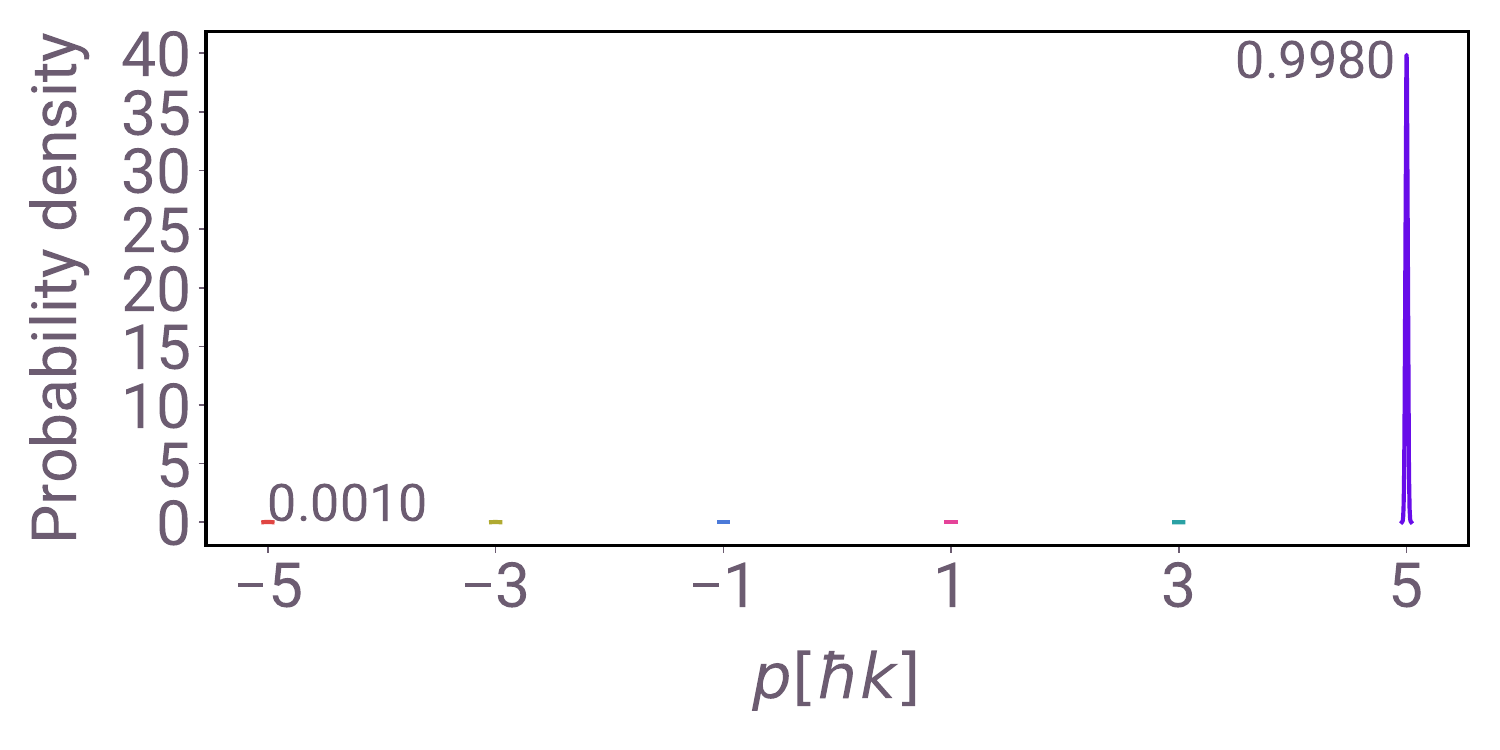}
\end{minipage}
\begin{minipage}[b]{0.32\textwidth}
\centering
\includegraphics[width=\textwidth]{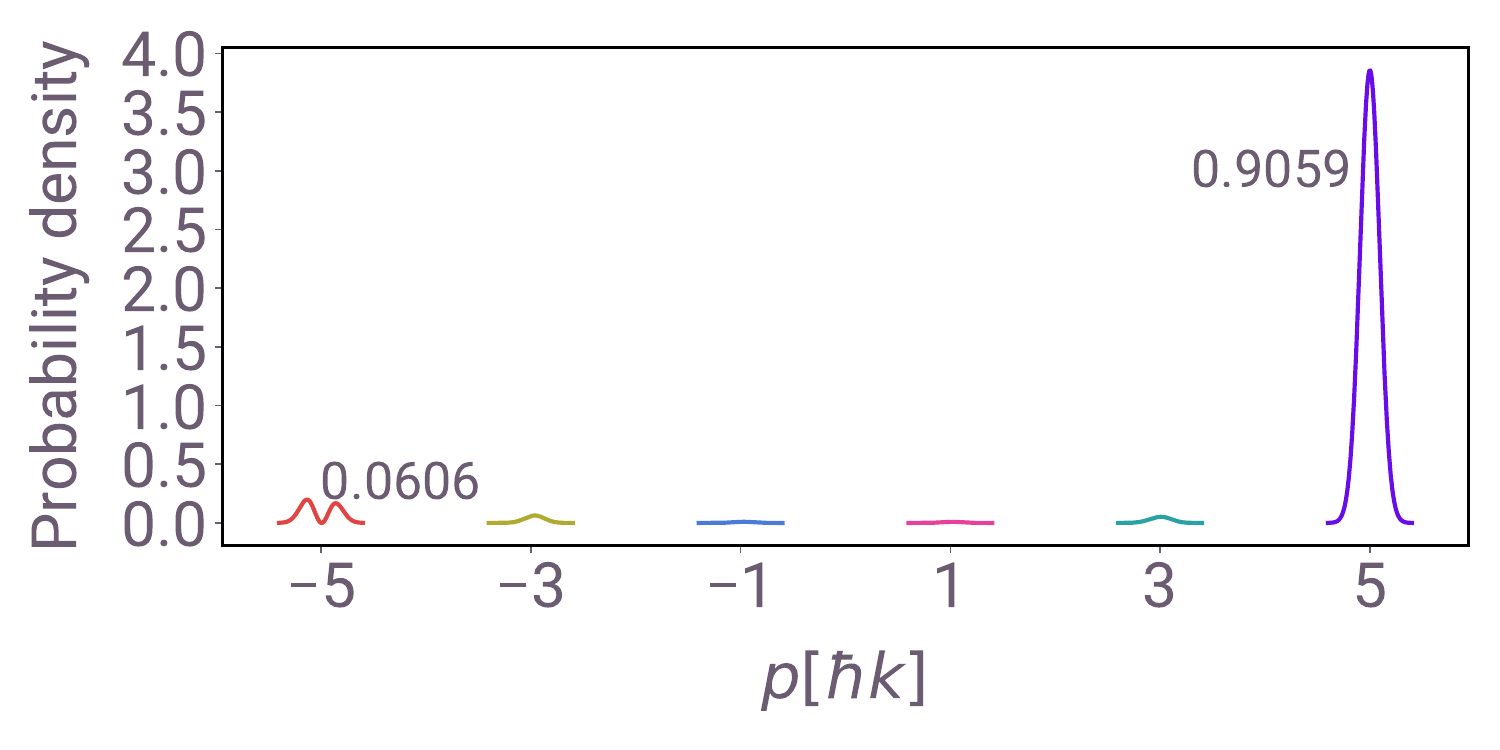}
\end{minipage}
\begin{minipage}[b]{0.32\textwidth}
\centering
\includegraphics[width=\textwidth]{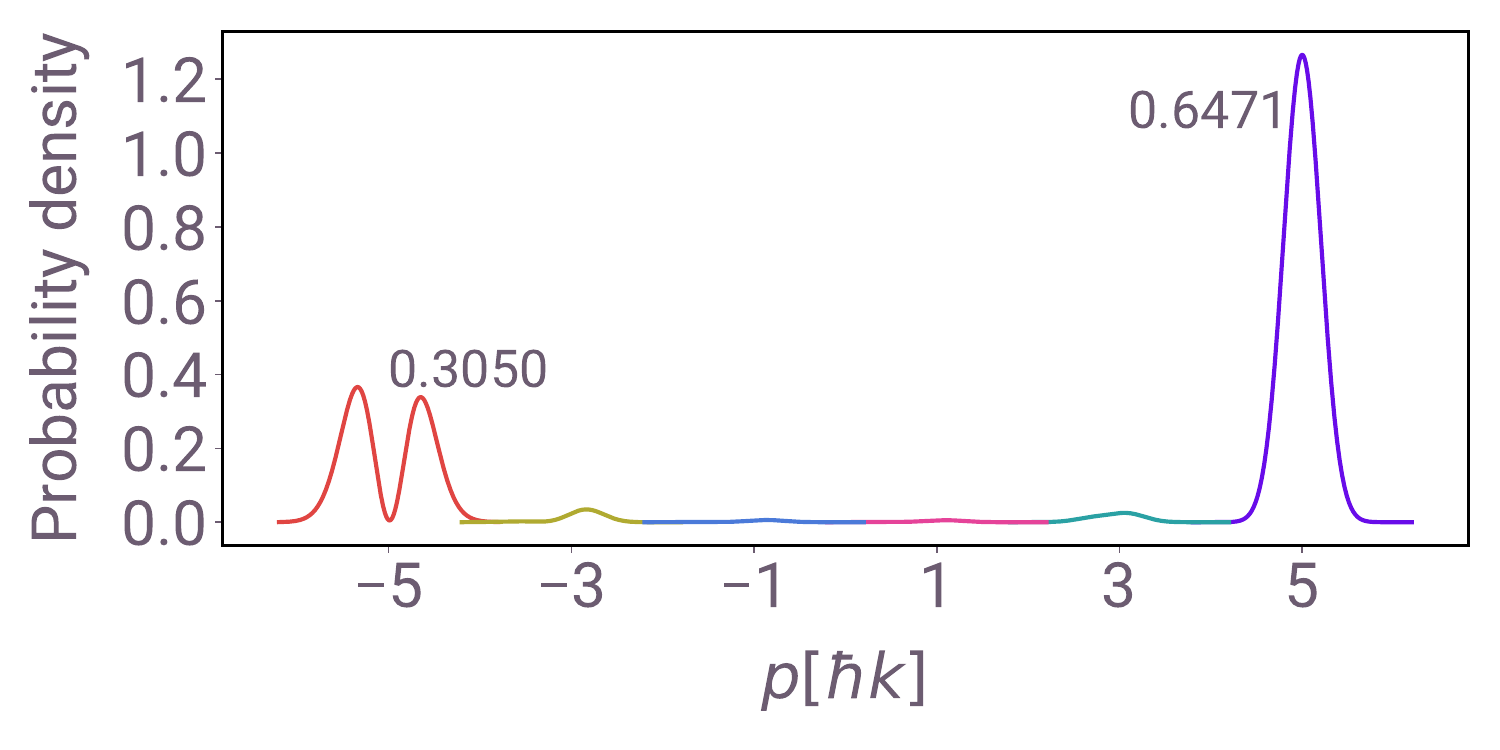}
\end{minipage}

\vspace{0.2cm}

% Second row: OCT pulse
\begin{minipage}[b]{0.32\textwidth}
\centering
\includegraphics[width=\textwidth]{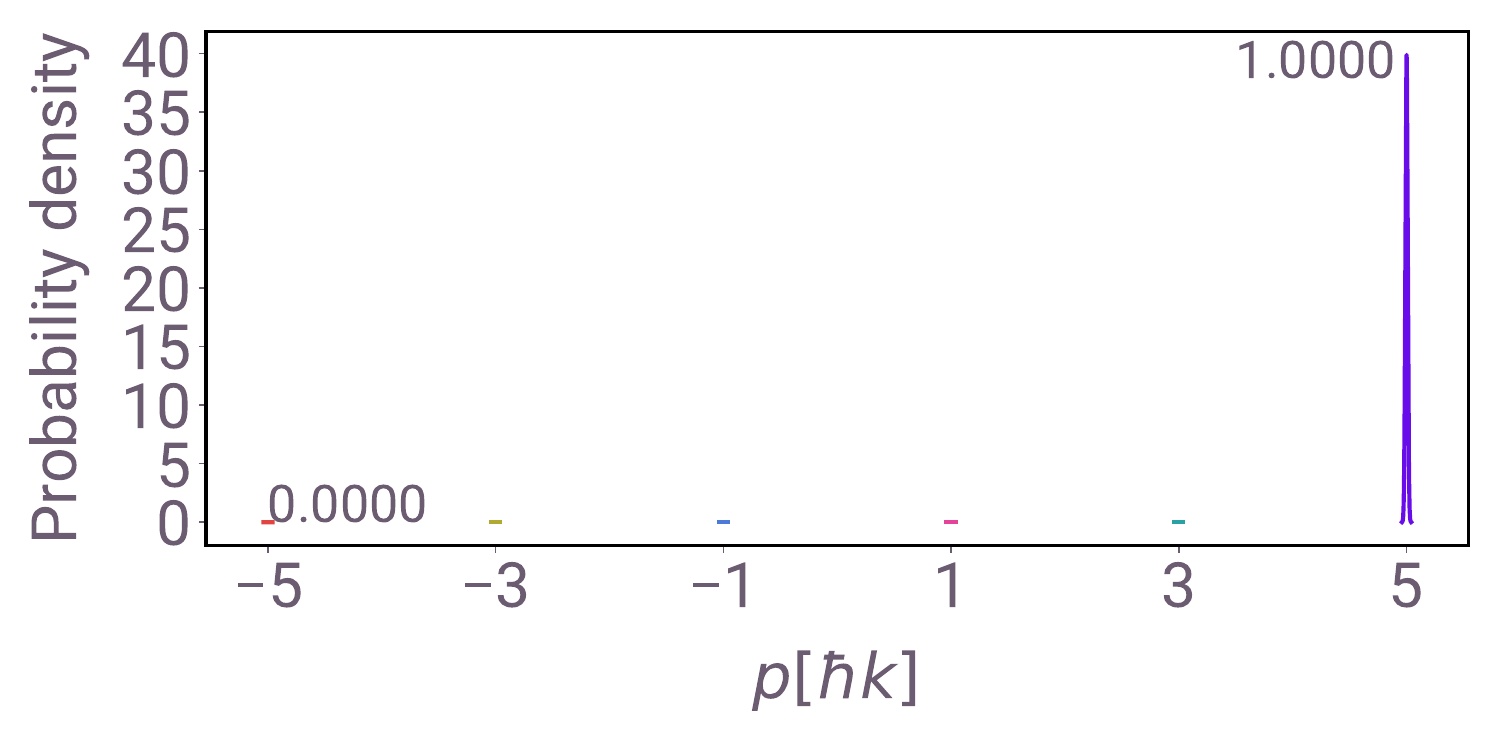}
\end{minipage}
\begin{minipage}[b]{0.32\textwidth}
\centering
\includegraphics[width=\textwidth]{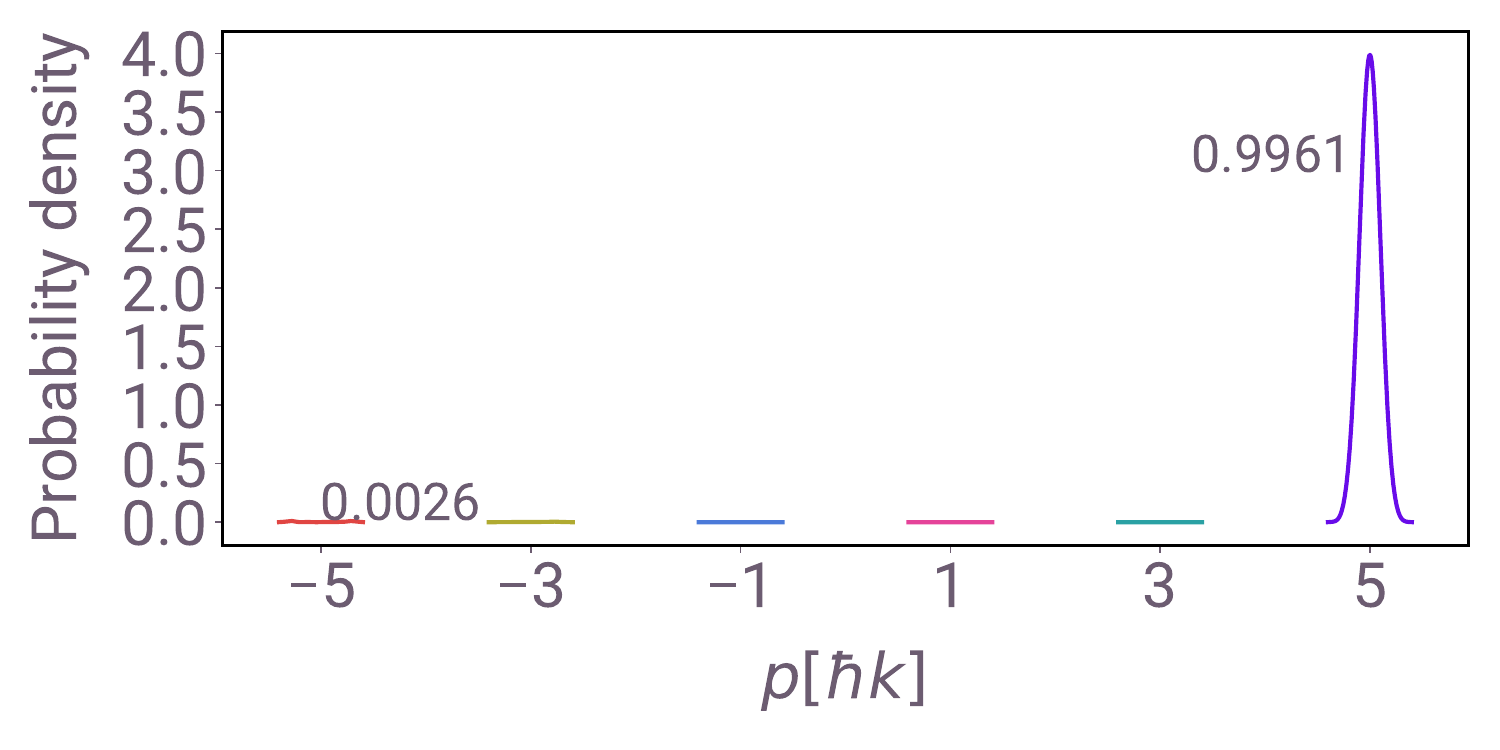}
\end{minipage}
\begin{minipage}[b]{0.32\textwidth}
\centering
\includegraphics[width=\textwidth]{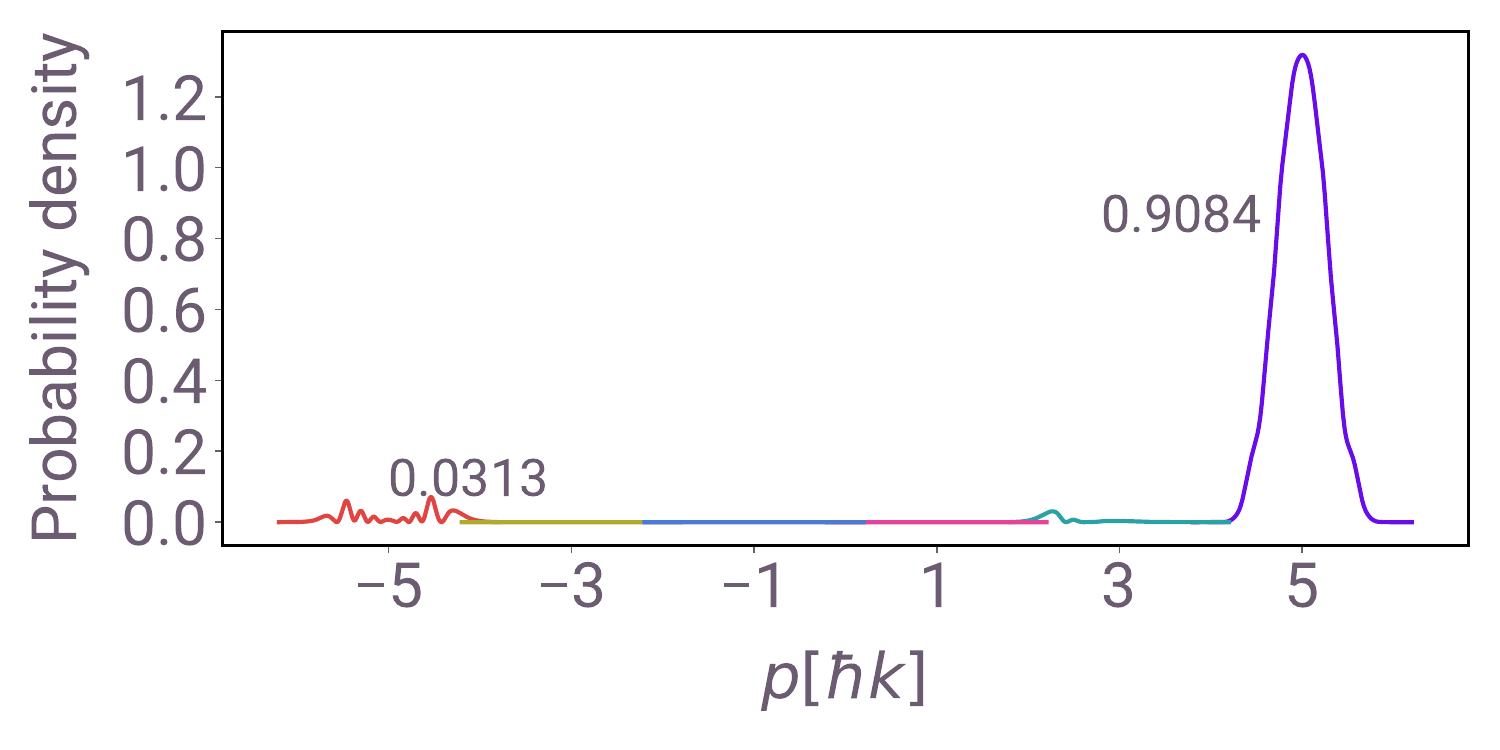}
\end{minipage}

\end{minipage}

\caption{Momentum distribution of an incoming state $\ket{-5 \hbar k}$ after 5th order Bragg mirror pulse. A Gaussian optimized pulse (top row) is compared to an OCT pulse (bottom row). The columns correspond to $ \SigmaGauss \in \{0.01,0.1,0.3\} \hbar k$. The population values of the main ports are indicated in the figures. The Gaussian pulse parameters are, in order, for the first row $\left(\Omega_0, \tau\right)=\{\left(33.69,0.382\right), \left(35.86,0.308\right),\left(36.94,0.292\right)\}$, where $\Omega_0$ is given in units of $\recoilfreq$ and $\tau$ in units of $\recoilfreq^{-1}$.}
\label{Fig:Mirrorn5}
\end{figure}

\begin{figure}[H]
\centering

% Left rotated label + main figure block
\begin{minipage}[c]{0.05\textwidth}
\rotatebox[origin=c]{90}{3rd order Bragg beam splitter}
\end{minipage}
\begin{minipage}[c]{0.93\textwidth}

% Column titles
\begin{minipage}[b]{0.32\textwidth}
\centering
\textbf{$\SigmaGauss = 0.01 \hbar k$}
\end{minipage}
\begin{minipage}[b]{0.32\textwidth}
\centering
\textbf{$\SigmaGauss = 0.1 \hbar k$}
\end{minipage}
\begin{minipage}[b]{0.32\textwidth}
\centering
\textbf{$\SigmaGauss = 0.3 \hbar k$}
\end{minipage}

\vspace{0.2cm}

% First row figures
\begin{minipage}[b]{0.32\textwidth}
\centering
\includegraphics[width=\textwidth]{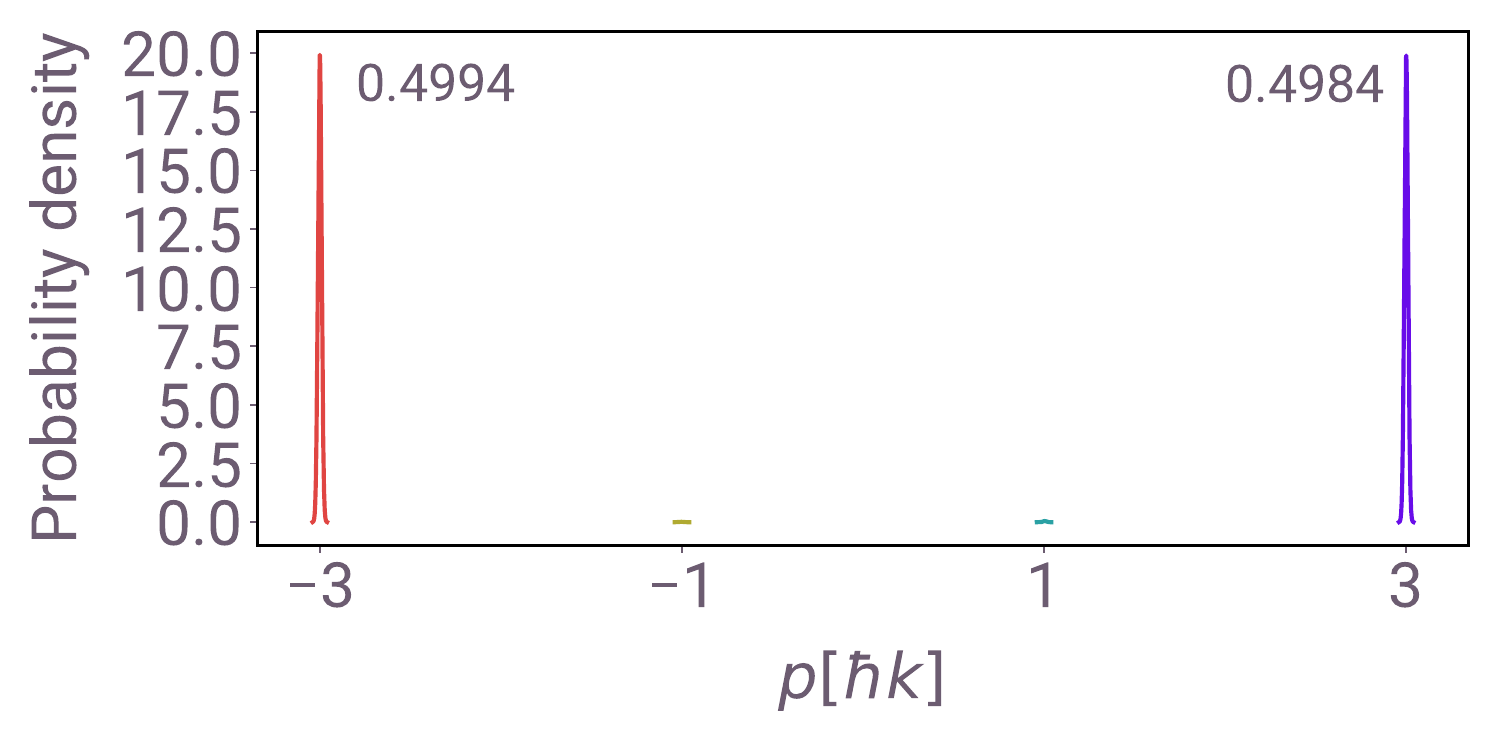}
\end{minipage}
\begin{minipage}[b]{0.32\textwidth}
\centering
\includegraphics[width=\textwidth]{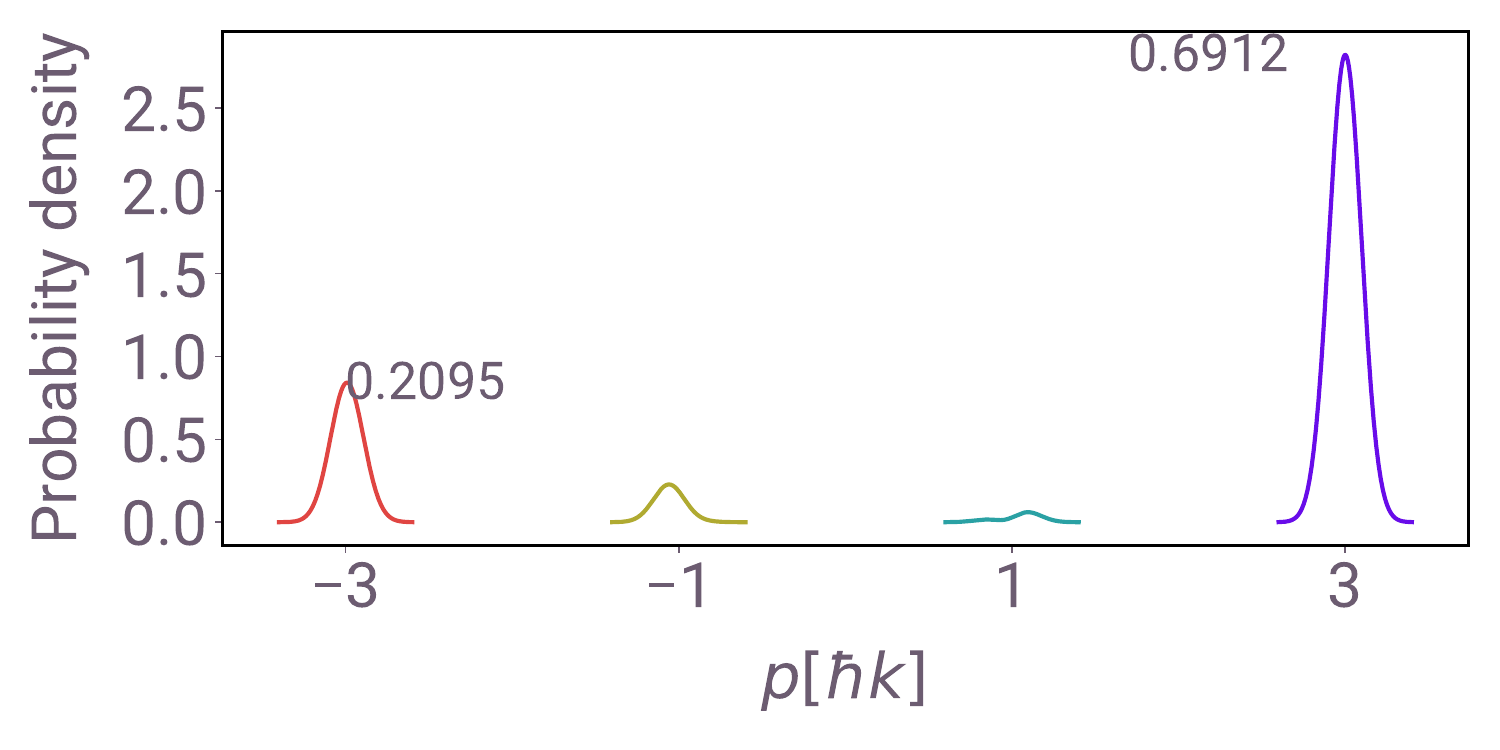}
\end{minipage}
\begin{minipage}[b]{0.32\textwidth}
\centering
\includegraphics[width=\textwidth]{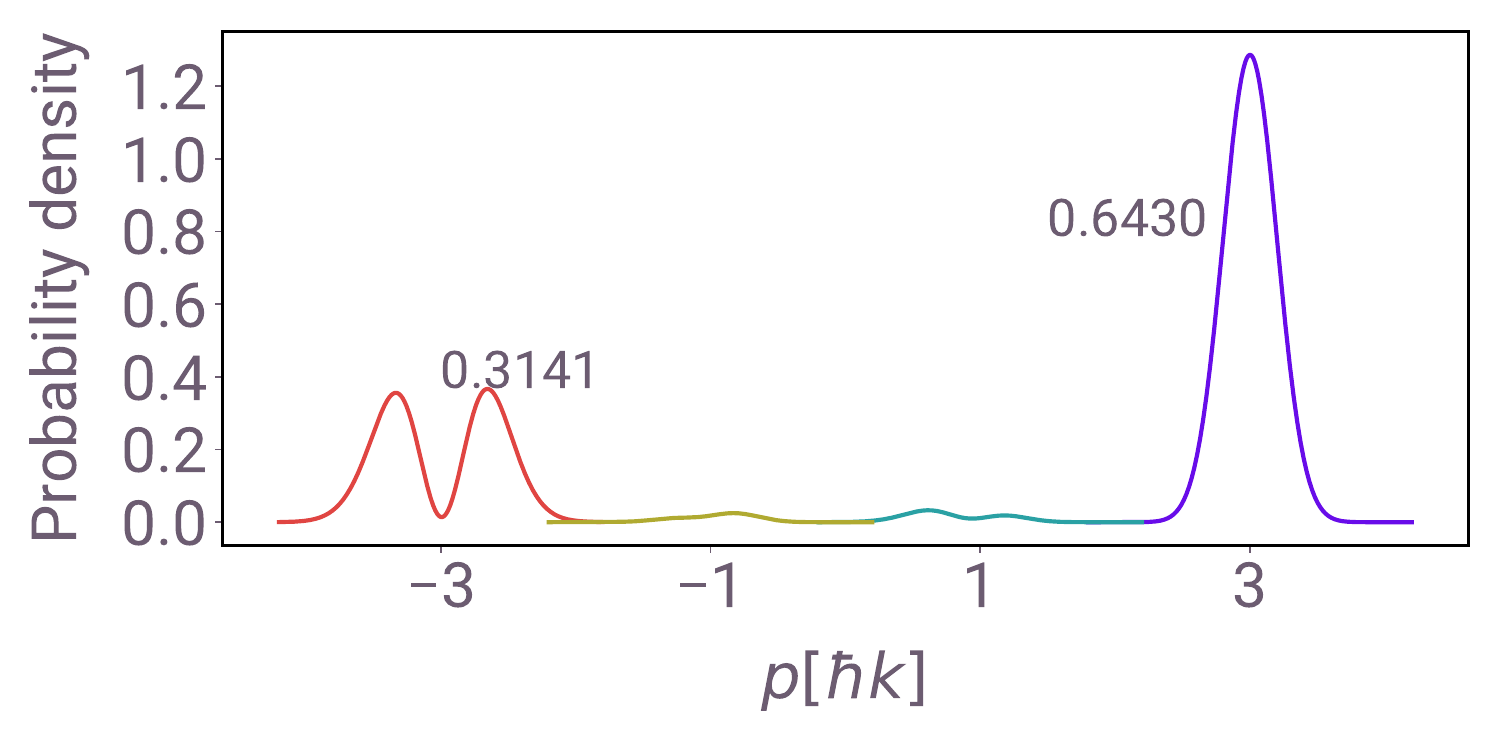}
\end{minipage}

\vspace{0.2cm}

% Second row figures
\begin{minipage}[b]{0.32\textwidth}
\centering
\includegraphics[width=\textwidth]{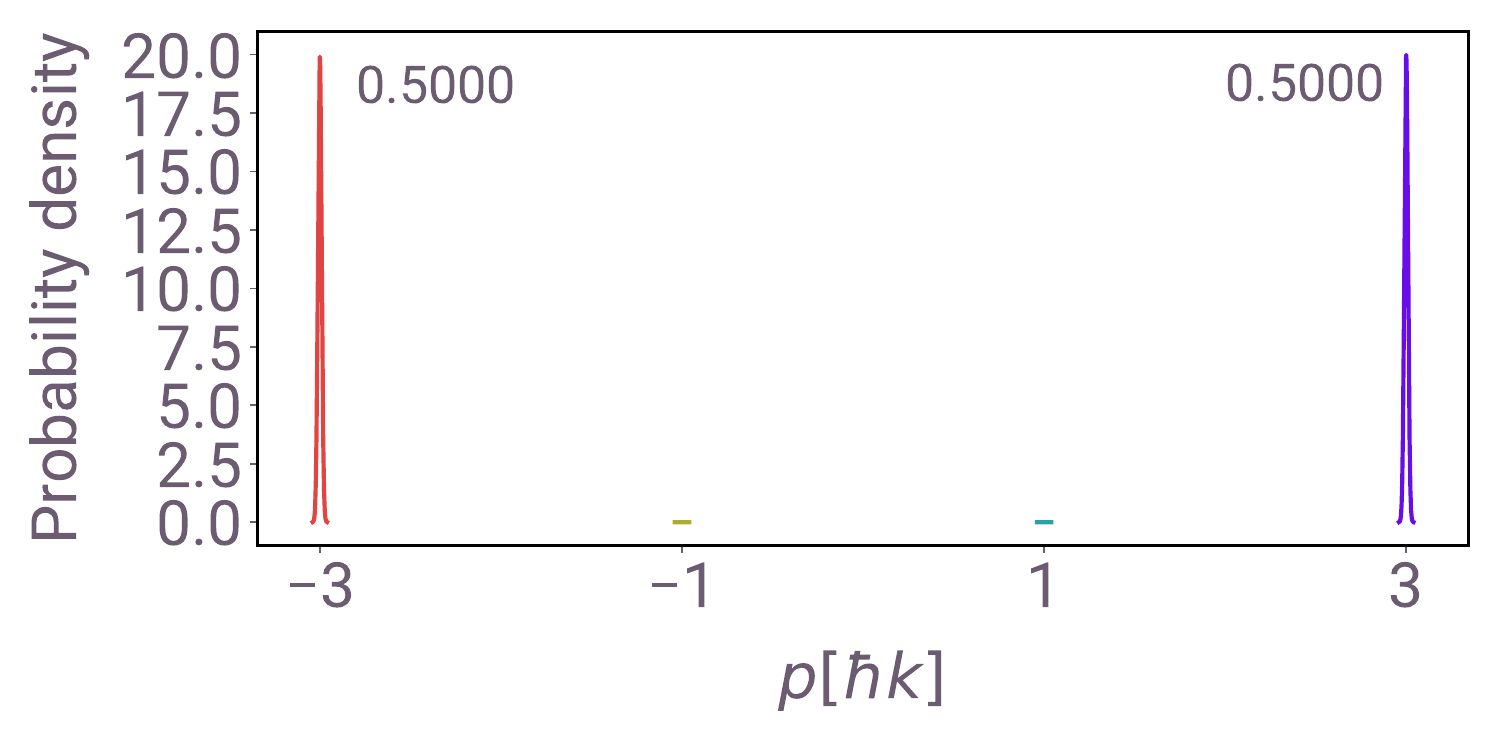}
\end{minipage}
\begin{minipage}[b]{0.32\textwidth}
\centering
\includegraphics[width=\textwidth]{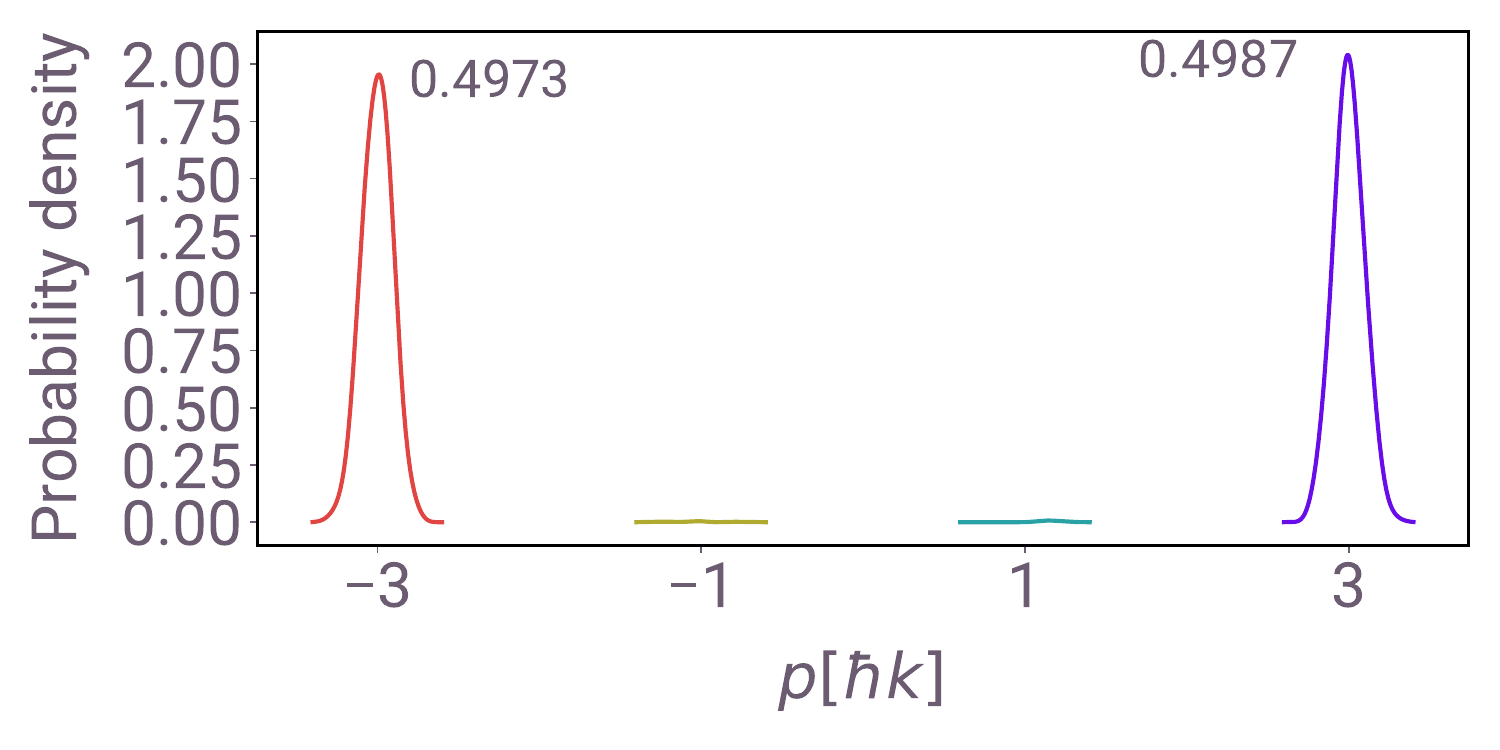}
\end{minipage}
\begin{minipage}[b]{0.32\textwidth}
\centering
\includegraphics[width=\textwidth]{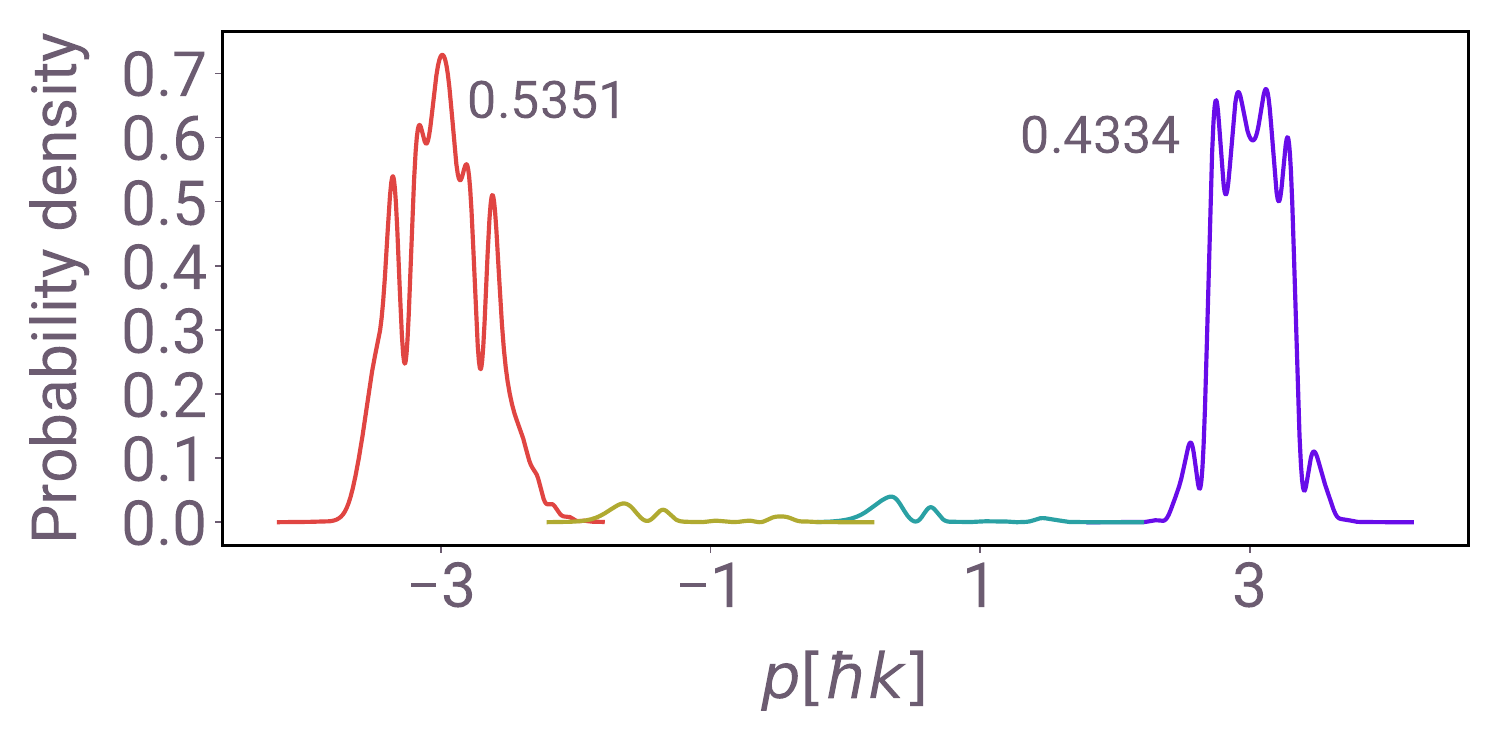}
\end{minipage}

\end{minipage}

\caption{Momentum distribution of an incoming state $\ket{-3 \hbar k}$ after 3rd order Bragg beam splitter pulse. A Gaussian optimized pulse (top row) is compared to an OCT pulse (bottom row). The columns correspond to $ \SigmaGauss \in \{0.01,0.1,0.3\} \hbar k$. The population values of the main ports are indicated in the figures. The Gaussian pulse parameters are, in order, for the first row $\left(\Omega_0, \tau\right)=\{\left(11.27,0.266\right), \left(15.59,0.178\right),\left(14.50,0.374\right)\}$, where $\Omega_0$ is given in units of $\recoilfreq$, and $\tau$ in units of $\recoilfreq^{-1}$.}
\label{Fig:BSn3}
\end{figure}

\begin{figure}[H]
\centering

\begin{minipage}[c]{0.05\textwidth}
\rotatebox[origin=c]{90}{5th order Bragg beam splitter}
\end{minipage}
\begin{minipage}[c]{0.93\textwidth}

% Column titles
\begin{minipage}[b]{0.32\textwidth}
\centering
\textbf{$\SigmaGauss = 0.01 \hbar k$}
\end{minipage}
\begin{minipage}[b]{0.32\textwidth}
\centering
\textbf{$\SigmaGauss = 0.1 \hbar k$}
\end{minipage}
\begin{minipage}[b]{0.32\textwidth}
\centering
\textbf{$\SigmaGauss = 0.3 \hbar k$}
\end{minipage}

\vspace{0.2cm}

% First row
\begin{minipage}[b]{0.32\textwidth}
\centering
\includegraphics[width=\textwidth]{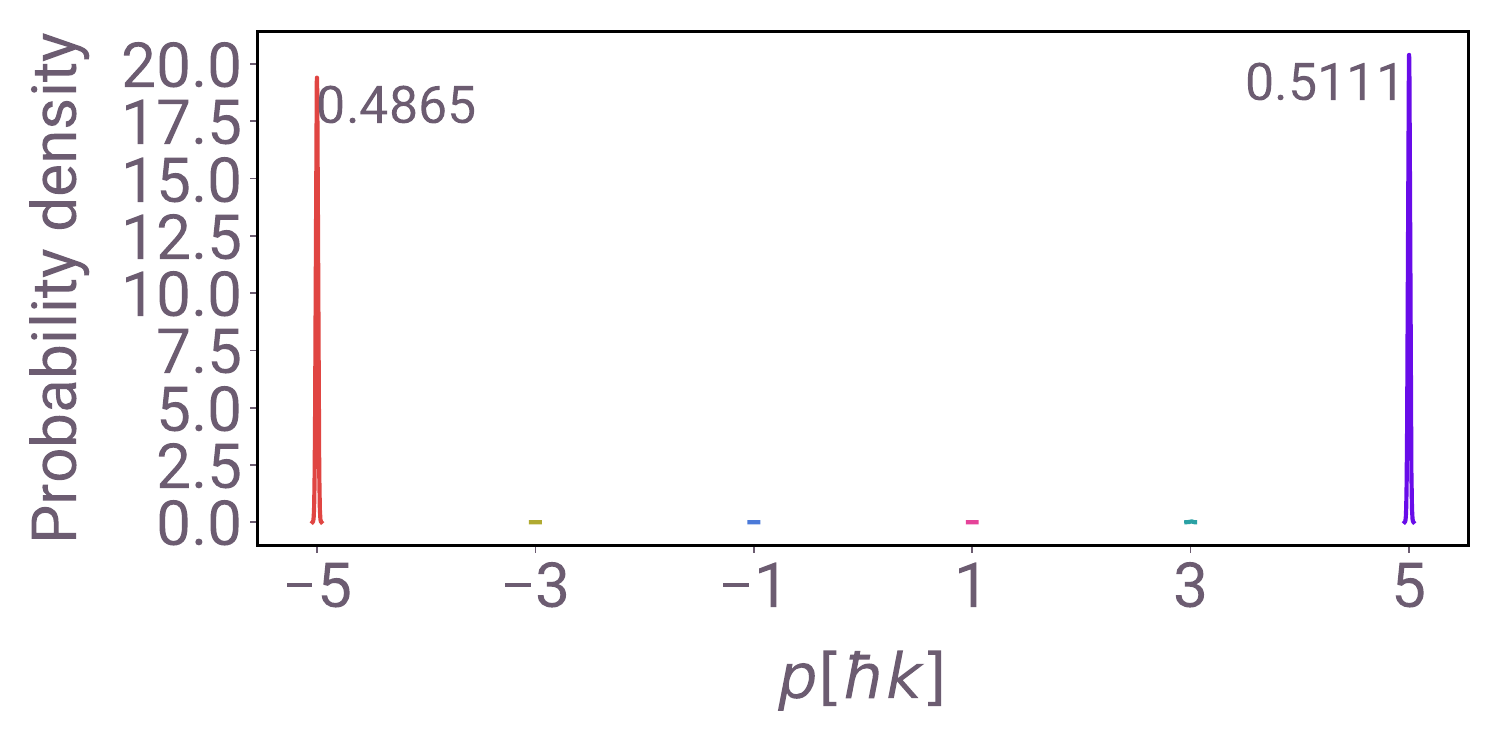}
\end{minipage}
\begin{minipage}[b]{0.32\textwidth}
\centering
\includegraphics[width=\textwidth]{Populations_after_interaction_GaussianBS,width_0.1,splitting_order_5.pdf}
\end{minipage}
\begin{minipage}[b]{0.32\textwidth}
\centering
\includegraphics[width=\textwidth]{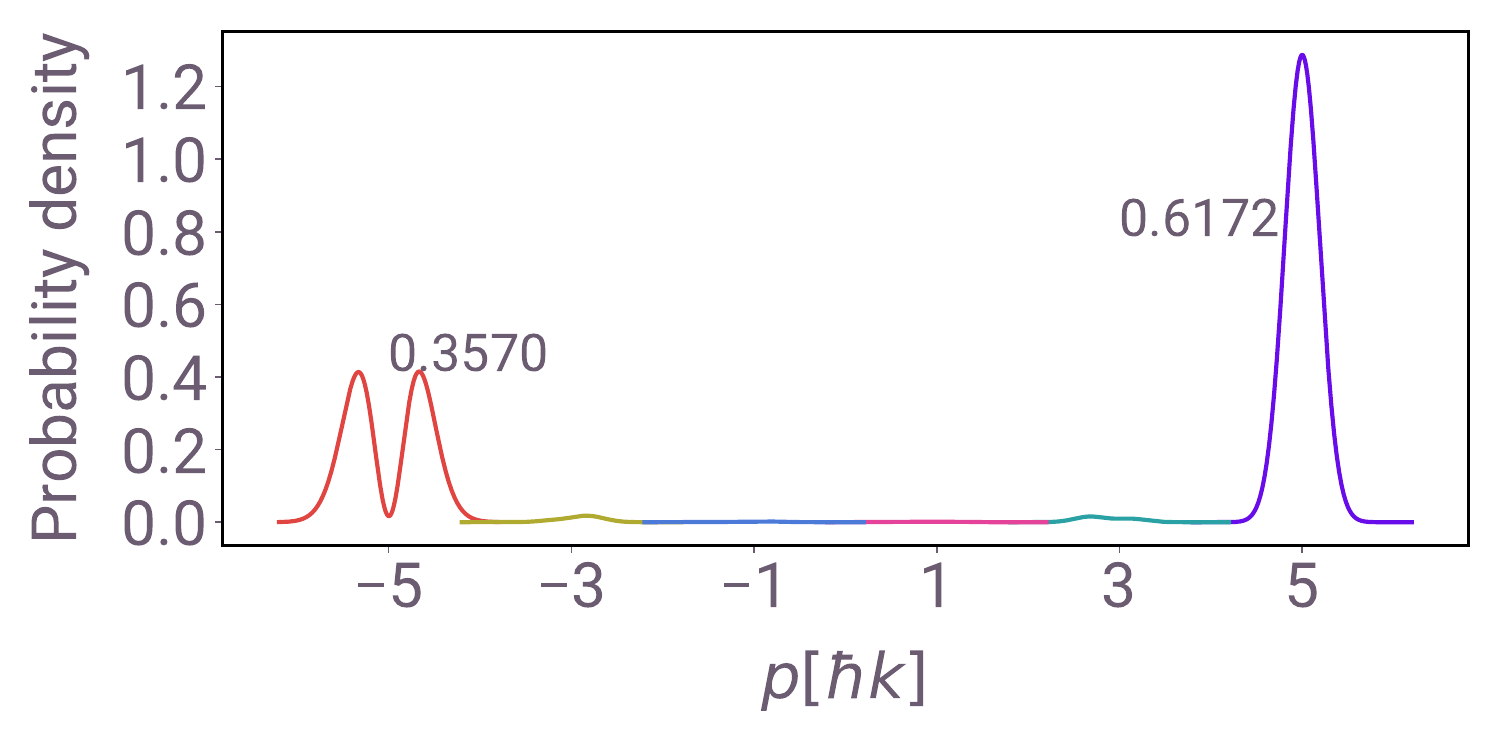}
\end{minipage}

\vspace{0.2cm}

% Second row
\begin{minipage}[b]{0.32\textwidth}
\centering
\includegraphics[width=\textwidth]{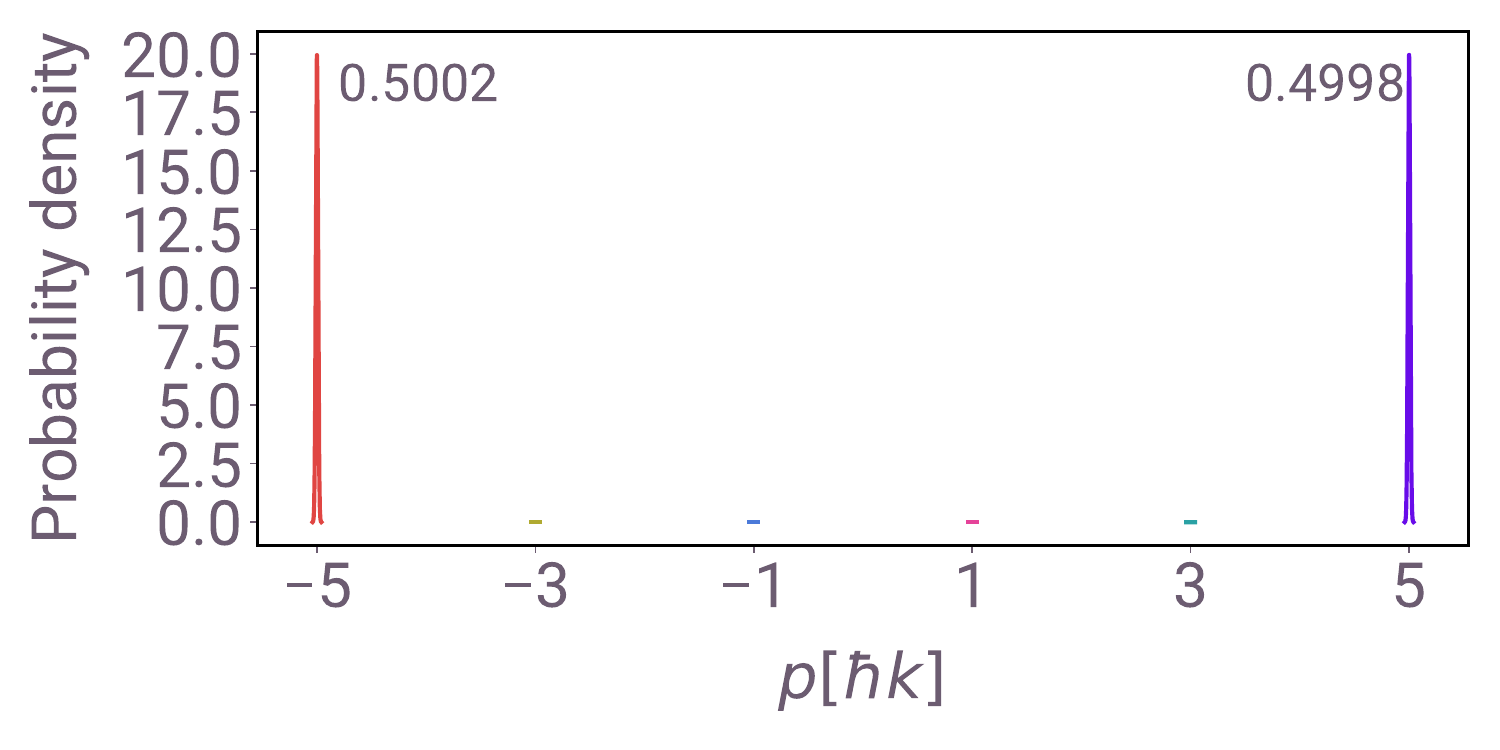}
\end{minipage}
\begin{minipage}[b]{0.32\textwidth}
\centering
\includegraphics[width=\textwidth]{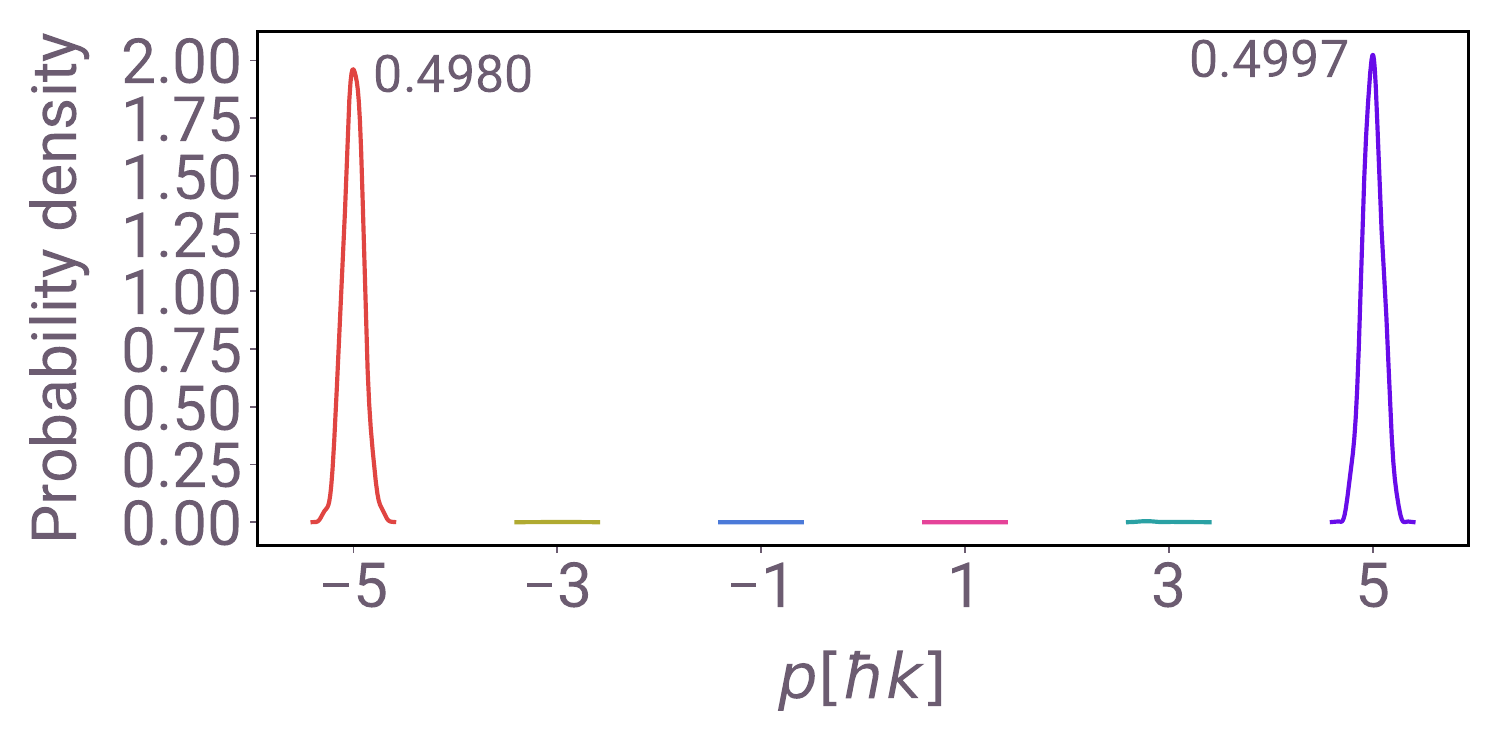}
\end{minipage}
\begin{minipage}[b]{0.32\textwidth}
\centering
\includegraphics[width=\textwidth]{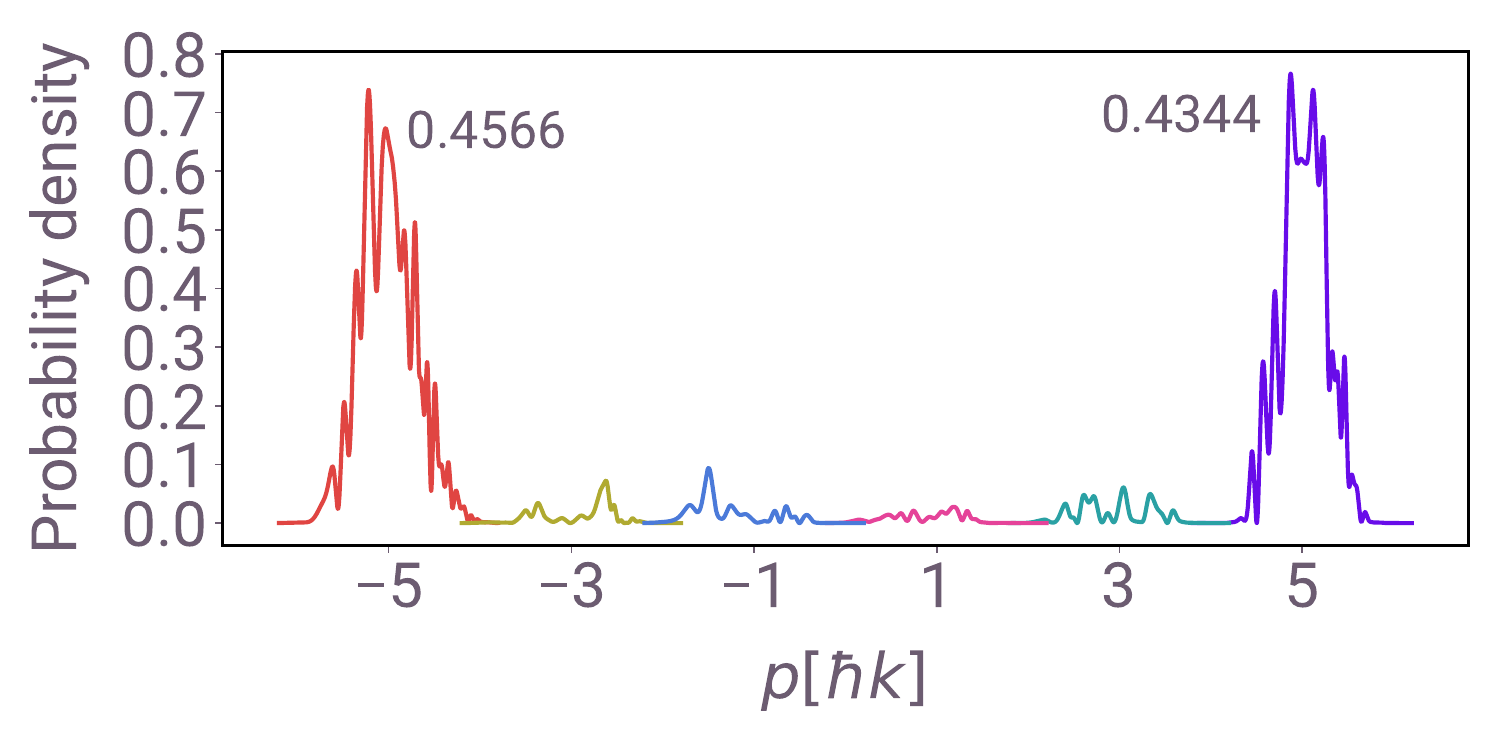}
\end{minipage}

\end{minipage}

\caption{Momentum distribution of an incoming state $\ket{-5 \hbar k}$ after 5th order Bragg beam splitter pulse. A Gaussian optimized pulse (top row) is compared to an OCT pulse (bottom row). The columns correspond to $ \SigmaGauss \in \{0.01,0.1,0.3\} \hbar k$. The population values of the main ports are indicated in the figures. The Gaussian pulse parameters are, in order, for the first row $\left(\Omega_0, \tau\right)=\{\left(31.06,0.211\right), \left(34,0.145\right),\left(34,0.336\right)\}$, where $\Omega_0$ is given in units of $\recoilfreq$ and $\tau$ in units of $\recoilfreq^{-1}$.}
\label{Fig:BSn5}
\end{figure}

\end{widetext}

%\clearpage

\end{document}